\documentclass[a4paper,manyauthors,nocleardouble,COMPASS]{cernphprep}

\RequirePackage[T1]{fontenc}

\RequirePackage{graphicx}
\RequirePackage{newtxtext}
\RequirePackage{newtxmath}
\RequirePackage{flushend}
\RequirePackage[numbers,sort&compress]{natbib}
\usepackage{grffile}
\usepackage{hyperref}
\hypersetup{
    colorlinks,
    citecolor=blue,
    filecolor=blue,
    linkcolor=blue,
    urlcolor=blue
}
\usepackage{cleveref}
\usepackage[disable]{todonotes}
\newcommand{\todoInl}[1]{}
\newcommand{\todoBl}[1]{}
\newcommand{\todoInlBl}[1]{}
\newcommand{\todoGr}[1]{}
\newcommand{\todoInlGr}[1]{}
\usepackage[final]{changes}
\usepackage[caption=false]{subfig}
\usepackage{booktabs}
\usepackage{xifthen}
\usepackage{caption}

\usepackage{packages/grubeMacros}
\usepackage{packages/formalismPWD}
\usepackage{packages/formalismRMF}
\usepackage{packages/dimensions}
\usepackage{packages/numbers}
\usepackage{packages/mpColorscheme}
\usepackage{packages/formalism_other}

\renewcommand*{\spectrumscale}{0.9}

\crefname{section}{Sec.}{Secs.}                             % set section name used by cleveref
\Crefname{section}{Section}{Sections}
\crefname{footnote}{footnote}{footnote}                     % set footnote name used by cleveref
\Crefname{footnote}{Footnote}{Footnotes}
\Crefname{figure}{Fig.}{Figs.}
\crefname{figure}{Fig.}{Figs.}
\Crefname{equation}{Eq.}{Eqs.}
\crefname{equation}{Eq.}{Eqs.}
\renewcommand{\thesection}{\Roman{section}}%
\renewcommand{\thesubsection}{\Alph{subsection}}%
\makeatletter
\renewcommand{\p@section}{}%
\renewcommand{\p@subsection}{\thesection\,}%
\renewcommand{\p@subsubsection}{\thesection\,\thesubsection\,}%
\makeatother

\newcommand{\seeEq}[1]{[see \cref{#1}]\xspace}

\graphicspath{{figures/}{../figures/}{./figures/logos}{../figures/plots/rmf/parameters/}{../figures/plots/prl/main/1+_0+_rho_770_0_K-_S/intensities/}{../figures/plots/prl/main/4+_1+_rho_770_0_K-_G/intensities/}{../figures/plots/prl/main/0-_0+_rho_770_0_K-_P/intensities/}{../figures/plots/prl/main/4+_1+_rho_770_0_K-_G/rhoAB/}{../figures/plots/prl/main/0-_0+_rho_770_0_K-_P/phases/}{figures/plots/rmf/parameters/}{figures/plots/prl/main/1+_0+_rho_770_0_K-_S/intensities/}{figures/plots/prl/main/4+_1+_rho_770_0_K-_G/intensities/}{figures/plots/prl/main/0-_0+_rho_770_0_K-_P/intensities/}{figures/plots/prl/main/4+_1+_rho_770_0_K-_G/rhoAB/}{figures/plots/prl/main/0-_0+_rho_770_0_K-_P/phases/}}

\newcommand{\ifequals}[3]{\ifthenelse{\equal{#1}{#2}}{#3}{}}
\newcommand{\case}[2]{#1#2} % dummy, ovewritten by new switch environment

\newcommand{\sub}[1]{\ensuremath{_{\mathrm{#1}}}}

\newcommand{\captionabove}[1]{\caption{#1}}

\newcommand{\captionLong}[2]{\caption[#1]{#1\xspace#2}}

\DeclareCaptionLabelFormat{subcaption-left}{\hspace*{2.5em}#2}
\DeclareCaptionLabelFormat{subcaption-lefter}{\hspace*{2.5em}#2}
\DeclareCaptionLabelFormat{subtable}{#2}
\newcommand{\supplementalsection}[1]{%
	\addtocounter{subsection}{1}%
	\subsection*{Supplemental Material \protect\thesubsection: #1}%
	\addcontentsline{toc}{subsection}{\protect\numberline{\thesubsection} #1}%
}

\usepackage{etex}
\usepackage{amsmath,amssymb,epsfig,fleqn,url,color,here,graphics,graphicx,rotating,multirow,wrapfig}
\usepackage{pbox}
\usepackage{dcolumn}
\usepackage{bigstrut}
\usepackage{soul,xcolor}
\usepackage{lineno, xcolor}

\usepackage[section]{placeins}
\usepackage{orcidlink}
\usepackage{tikz}

\begin{document}
%\linenumbers
\begin{titlepage}
\PHnumber{2025-086}
\PHdate{\today}
\title{\mytitle}

\Collaboration{The COMPASS Collaboration}
\ShortAuthor{The COMPASS Collaboration}
%\documentclass[aps,prl,preprint,superscriptaddress]{revtex4-2}
%\documentclass[aps,prl,reprint,groupedaddress]{revtex4-2}

% You should use BibTeX and apsrev.bst for references
% Choosing a journal automatically selects the correct APS
% BibTeX style file (bst file), so only uncomment the line
% below if necessary.
\bibliographystyle{apsrev4-2}

\begin{abstract}
	% 600 CHARACTERS
We measured the strange-meson spectrum in the scattering reaction \reactionK with the COMPASS spectrometer at CERN.
Using the world's largest sample of this reaction, we performed a comprehensive partial-wave analysis of the mesonic final state.
It substantially extends the strange-meson spectrum covering twelve states with masses up to \SI{2.4}{\GeVcc}. We observe the first candidate for a crypto-exotic strange meson with $\JP=\zeroM$ and find \PKThree* and \PKFour* states consistent with predictions for the ground states.

\end{abstract}
	
\vfill
\Submitted{(to be submitted to Phys. Rev. Letters)}
\end{titlepage}
{
\pagestyle{empty}
\clearpage
}
%\tableofcontents
\clearpage

\setcounter{page}{1}

%=============================================================================
%=============================================================================
%=============================================================================
\section{Introduction}

\begin{figure}[tb]%
	\centering%
	\includegraphics[width=\spectrumscale\linewidth]{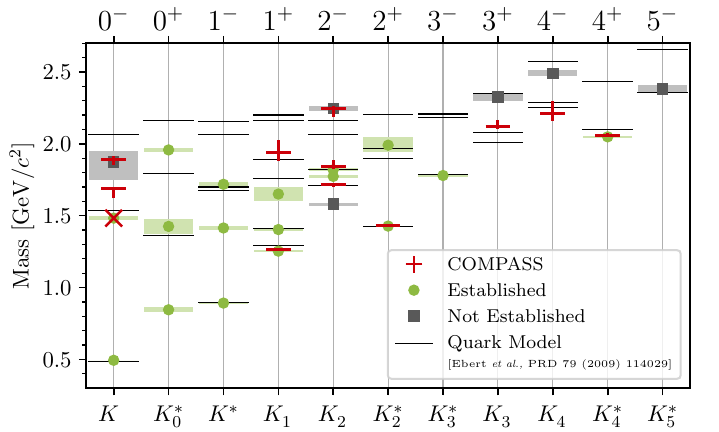}%
	\caption[Mass spectrum of strange mesons spectrum.]{Mass spectrum of strange mesons grouped by \JP quantum numbers (columns). Red horizontal lines represent our measurements, where the vertical lines indicate the total uncertainties on the mass. The red crosses represents our observation of the \PK[1460], whose mass and width parameters were fixed to the LHCb measurement in \refCite{LHCb:2017swu}. Green points show established states; gray squares not established ones~\cite{ParticleDataGroup:2024cfk}. The box heights indicate the mass uncertainties. Black horizontal lines represent the quark-model calculation from \refCite{Ebert2009}. %As we show only masses below \SI{2.7}{\GeVcc} for a better visualization, the not-established \PK[3100] is omitted here.
	}%
	%	\caption[Spectrum of strange mesons as obtained from our analysis.]{Spectrum of strange mesons, \ie nominal masses of strange mesons grouped by their \JP quantum numbers. The red crosses represented our measured mass values. The green data points show established states, the gray data points not established states as listed by the PDG~\cite{ParticleDataGroup:2024cfk}. The similarly colored boxes represent the corresponding mass uncertainties. The black horizontal lines show the masses of states as predicted by the quark-model calculation in \refCite{Ebert2009}. As we show only masses below \SI{2.7}{\GeVcc} for a better visualization, the not-established \PK[3100] is omitted here.}%
	\label{fig:kaonSpectrum}%
\end{figure}

Our knowledge and understanding of strongly bound systems of quarks and anti-quarks is mostly limited to non-strange light and to heavy hadrons. Strange mesons are only poorly known and dedicated theoretical studies are limited, but are necessary to bridge the gap between light and heavy hadrons.
At present, 17 strange mesons are experimentally established (see PDG~\cite{ParticleDataGroup:2024cfk} and \cref{fig:kaonSpectrum}).\nocite{LHCb:2017swu}
Eight other states need confirmation, most of them observed by only a single experiment and in only a single decay channel. 
%The states, such as the \PKTwo[2250]~\cite{Baubillier1981,Cleland:1980ya,Bari-Birmingham-CERN-Milan-Paris-Pavia:1983nwf,Lissauer1970,Chliapnikov1979}, the \PKThree[2320]~\cite{Bari-Birmingham-CERN-Milan-Paris-Pavia:1983nwf}, and the \PKFour[2500]~\cite{Cleland:1980ya}, are observed  by a single, rarely several experiments and in one final state only. 
%States, such as \PKTwo[2250]~\cite{Baubillier1981,Cleland:1980ya,Bari-Birmingham-CERN-Milan-Paris-Pavia:1983nwf,Lissauer1970,Chliapnikov1979}, \PKThree[2320]~\cite{Bari-Birmingham-CERN-Milan-Paris-Pavia:1983nwf}, and \PKFour[2500]~\cite{Cleland:1980ya}, are observed  by a single, rarely several experiments and in one final state only. 
Models using constituent quarks and SU(3) flavor symmetry provide a pattern of \qqbarPrime states~\cite{Ebert2009,Oudichhya:2023lva,Taboada-Nieto:2022igy,Pang:2017dlw,Godfrey1985} (black horizontal lines in \cref{fig:kaonSpectrum}).
However, many of the predicted states are still unobserved.
They are challenging to identify experimentally due to the dense strange-meson spectrum, \eg caused by the fact that $K_J$ states appear as pairs close in mass, arising from the singlet and triplet \qqbarPrime states.

In recent years, the study of meson spectra has revealed many exotic phenomena beyond \qqbarPrime states. For example, four-quark states containing heavy quarks have been discovered, \eg~\refCite{LHCb:2021uow}, %\cite{Choi2003,Aaltonen2009,Ablikim2013,LHCb:2016axx,Aaij2022}
and \PpiOne, a candidate for a light hybrid meson, has been firmly established~\cite{COMPASS:2021ogp}.
There is no evidence yet for strange counterparts, but the \PKZeroSt[700]/$\kappa$ that is discussed as a four-quark state~\cite{ParticleDataGroup:2022pth_rev_strange_mesons}.
While some exotic non-strange light and heavy mesons are clearly identified by their non-\qqbarPrime quantum numbers, in the strange sector no such model-independent observable exists that distinguishes quark-model from exotic states.
Therefore, strange exotics can only be observed as additional states within the already dense spectrum of predicted \qqbarPrime states. Consequently, these states are called supernumerary or crypto-exotic.
To establish such states, it is necessary to map out the strange-meson excitation spectrum over a wide mass range.

In this paper, we present the strange-meson spectrum observed in the diffractive scattering reaction \reactionK with the COMPASS spectrometer at CERN. We performed a partial-wave analysis, from which we obtained the masses and widths of eleven strange-meson resonances with masses up to \SI{2.4}{\GeVcc}. Details of this analysis can be found in \refCite{Wallner:2022prx}.

%=============================================================================
%=============================================================================
%=============================================================================
\section{Method}
\label{sec:pwa}

The \SI{190}{\GeVc} M2 beam at CERN contains about \SI{2}{\percent} \PKNeg, which were identified by two differential Cherenkov counters. The \Kpipi final-state particles were measured with the COMPASS magnetic spectrometer and identified with a ring-imaging Cherenkov detector~\cite{Abbon:2014aex}. We imposed energy-momentum conservation and required that the reconstructed scattering vertex formed by all particles is inside the target volume.
In total, our \Kpipi sample consists of \num{720000} exclusive events, exceeding the best previous measurement~\cite{ACCMOR:1981yww} by a factor of 3.6. 

We performed a partial-wave analysis in two stages to identify spin and parity of strange mesons decaying into \Kpipi and to extract their masses and widths analogously to our work in \mbox{\refsCite{COMPASS:2015gxz,COMPASS:2018uzl}}. 
First, we determined the contributions of partial waves to the measured spectrum in terms of the spin-density matrix \sdm*.
To avoid making assumptions about the resonance content of the \Kpipi system at this stage, we independently determined $\rho_{ab}(\mKpipi, \tpr)$ in narrow bins of \mKpipi and four bins of the reduced four-momentum transfer squared \tpr between the beam kaon and the target proton in the ranges of \SIvalRange{1.0}{\mKpipi}{3.0}{\GeVcc} and \SIvalRange{0.1}{\tpr}{1.0}{\GeVcsq}.
To this end, the distribution of \Kpipi events in the phase-space variables $\tau$ in an \mKpipiTpr bin was modeled as %\footnote{ %
	%We allow all combinations within the following loose constraints: $J\leq 7$; $L\leq 7$; positive naturality of the exchange particle; and twelve isobars: two $K\pi$ $S$-wave amplitudes, $K^*(892)$, $K^*(1680)$, $K^*_2(1430)$, $K^*_3(1780)$, a broad $\pi\pi$ $S$-wave amplitude, $f_0(980)$, $f_0(1500)$, $\rho(770)$, $f_2(1270)$, and $\rho_3(1690)$.
	% Also, a so-called flat wave, which has a uniform distribution in $\tau$, is incoherently added to the intensity model. Since the flat wave does not pick up significant intensity, we drop it in \cref{eq:intensity} for simplicity.}\todo{Drop comment about flat wave as formally, we can also represent it as part of \sdm*?}
\begin{equation}
	\mathcal{I}(\tau) = \sum\limits_{a,b}^{\text{waves}} \psi_a(\tau) \rho_{ab}\, \psi_b^*(\tau). \label{eq:intensity}
\end{equation}
Partial waves are labeled by $a = \WaveK{J}{P}{M}{\varepsilon}{\isobar*}{\xi}{L}$, where $J^P$ are spin and parity, and $M^\varepsilon$ is the spin projection along the beam axis expressed in the reflectivity basis~\cite{Chung1975}.
We assume that the \Kpipi final state results from subsequent two-body decays via an intermediate isobar \isobar*, \ie a two-body resonance. $L$ is the orbital angular momentum between the bachelor particle $\xi$, which is a pion or a kaon, and the isobar.
The decay amplitudes $\psi_a$ are computed using the isobar model (see \refsCite{Hansen:1973gb,Herndon:1973yn,Ketzer:2019wmd}).
To infer the waves included in $\sum_{a,b}^{\text{waves}}$ that contribute significantly to the data, we use regularization-based model-selection techniques described in \refCite{Wallner:2022prx} to reduce the large pool of 596 allowed waves constructed using the loose constraints $J\leq7$; $L\leq 7$; $M=0,1,2$; $\varepsilon=+$; and a total of twelve isobar resonances in the $\PKNeg\PpiPos$ and $\PpiNeg\PpiPos$ subsystems listed in Tab.~5.2 of \refCite{Wallner:2022prx}.
Isobar resonances are parameterized by relativistic Breit-Wigner amplitudes~\cite{PhysRev.49.519}, but special parameterizations from \refsCite{IgorA.Kachaev2004,Au:1986vs,COMPASS:2015gxz} and \refCite{Palano:2017nex} are used for low-mass $\pi\pi$ and $K\pi$ $S$-wave isobars, respectively.
We obtain $\rho_{ab}(\mKpipi, \tpr)$ by performing unbinned extended maximum-likelihood fits of \cref{eq:intensity} to the data, accounting for experimental acceptance.
The statistical uncertainties are estimated by Bootstrapping~\cite{Efron1979}, \ie by randomly resampling and refitting the data sample.

Besides signal events, several backgrounds from other reactions, \eg from \reaction caused by misidentified pions, leak into the \Kpipi sample.
All backgrounds are accounted for by parameterizing $\rho_{ab}$ as a matrix of rank three~\cite{Wallner:2022prx}. This effectively decomposes their intensity distribution into the \Kpipi partial-wave basis and adds them incoherently in \cref{eq:intensity}.

In the second stage, we performed a resonance-model fit (\rmf) to the measured \sdm, to extract resonance signals from the partial waves and to determine their masses and widths.
%In the second stage (resonance-model fit - \rmf), we extracted the resonance signals from the partial waves and measured their masses and widths.
%To this end, $\rho_{ab}\mKpipiTpr$ for 14 selected partial waves (see table ?? in the supplemental material\todoGr{Add correct table numer}) is modeled as an incoherent sum over true \Kpipi and two incoherent background contributions (see supplemental material for details):
We selected 14 partial waves with signals covering almost all \JP in \cref{fig:kaonSpectrum} (Table~II in the supplemental material~\cite{supplemental}\todoGr{Check reference to supplemental material}) and modeled the corresponding $\rho_{ab}\mKpipiTpr$ as a sum over true \Kpipi and two incoherent background contributions, \ie
\begin{equation}
	\rho_{ab} =  \transitionAmpl*[\waveLabelA][] \transitionAmpl*[\waveLabelB][*] + \rho_{ab}^{3\pi} + \rho_{ab}^{\text{rem bkg}},\label{eq:rmf}
\end{equation}
where $\rho_{ab}^{3\pi}$ is a fixed parameterization for the largest incoherent background from \reaction, which we determined from our own high-precision \threePi data~\cite{COMPASS:2015gxz}, and $\rho_{ab}^{\text{rem bkg}}$ effectively subsumes other remaining incoherent backgrounds by an adapted resonanceless parameterization (see \refCite{supplemental} for details).

The true \Kpipi amplitudes
%\todo{@Stephan: what yo you mean with signal?\\Stefan: The same as in the rest of the text, \Kpipi. Is this unclear here?} 
are modeled as
%\begin{equation}
%	\rho_{ab}^{K\pi\pi}\mKpipiTpr =  \mathcal T_a\mKpipiTpr \mathcal T_b^*\mKpipiTpr,
%\end{equation}
\begin{equation}
	\transitionAmpl[\waveLabelA][] = \mathcal{K}\mKpipiTpr \sum\limits_{\componentLabelK} \coupling[\waveLabelA][\componentLabelK][]\, \dynamicF*(\mKpipi),
\end{equation}
which is a coherent sum over all model components \componentLabelK for the wave \waveLabelA. These include resonances and non-resonant components that account for multi-Regge processes \cite{Deck:1964hm,JPACCollaboration2021}.
The dynamic amplitudes $\mathcal D_k(\mKpipi)$ of the resonance components are parameterized by relativistic Breit-Wigner functions, those of the non-resonant components by an empirical real-valued function (see \refCite{supplemental}).
%, with an intensity decreasing exponentially with \mKpipi
\coupling[\waveLabelA][\componentLabelK][] is the coupling amplitude of component $k$ within the wave $a$, and is an independent constant for each \tpr bin.
$\mathcal K(\mKpipi, \tpr)$ accounts for the phase-space volume and encodes the mass-dependent production by Pomeron exchange~\cite{Atayan1991}. %~\cite{Atayan1991,Adloff:1997sc,Abe:1993xx}.

%Using a $\chi^2$ fit, the 
%We fit the resonance model to the spin-density matrix elements of the 14 selected partial waves, with \coupling[\waveLabelA][\componentLabelK][] and the shape parameters of $D_k(\mKpipi)$, \eg the nominal masses and widths of the resonances, being free parameters.

The \coupling[\waveLabelA][\componentLabelK][] and shape parameters of $\dynamicF*(\mKpipi)$, including \eg the masses and widths, were determined by a \chis fit to the spin-density matrix elements of the 14 waves, taking into account their statistical correlations. These elements comprise 14 partial-wave intensities and 91 complex-valued interference terms.
We performed extensive studies in order to estimate the systematic uncertainties of measured masses and widths (see section~B\todoGr{Check supplemental reference} in \refCite{supplemental} for details).

%=============================================================================
%=============================================================================
%=============================================================================
\section{Ambiguous Identification of Final-State Particles}
\label{sec::constraints}

At COMPASS, the identification of final-state kaons and pions is limited to momenta below about \SI{50}{\GeVc}. Events for which the identification was thus ambiguous were discarded from the analysis.
This ambiguous identification renders a subset of partial waves indistinguishable for masses below about \SI{1.6}{\GeVcc}. They are therefore excluded from physics analysis.
However, the 14 waves used in the resonance-model fit are interpretable, allowing us to extract strange mesons from them.
This was confirmed in several systematic studies including extensive Monte Carlo input-output comparisons.

%Our We were limited in the selection of waves that entered the resonance-model fit by limitations intrinsic to this analysis.
%Our final-state particle identification separates kaons from pions only up to momenta of about \SI{50}{\GeVc}, while they have momenta up \SI{190}{\GeVc}.
%Therefore, we cannot perfectly suppress incoherent backgrounds, \eg from the reaction \reaction, and must treat them in the analysis as described above.
%Signal events where the \PKNeg and the \PpiNeg have momenta above about \SI{50}{\GeVc} cannot be used in the analysis, resulting in zero acceptance in phase-space regions. 
%This reduces the distinguishability of certain partial waves leading to large analysis artifacts in these waves. Consequently, the results for affected waves cannot be used for physics analysis. However, studies of the partial-wave properties revealed that only a subset of waves is affected, while the remaining waves are robust and can be interpreted in terms of physics signals, such as the 14 waves selected for the resonance-model fit.
%We validated this in comprehensive Monte Carlo input-output studies.

%=============================================================================
%=============================================================================
%=============================================================================
\section{Results}
\label{sec:results}

\begin{table}%
	\setTableStretch{1.2}%
	\centering%
	\caption{Our resonance parameters. The first quoted uncertainties are statistical, the second systematic.}%
	\label{tab:parameters}%
	\begin{tabular}{l|ll}
		Resonance    & $m_0$ [\si\MeVcc] & $\Gamma_0$ [\si\MeVcc] \\ \midrule
		\PK[1690]    & \MtK*{1630}       & \GtK*{1630}            \\
		\PK[1830]    & \MtK*{1830}       & \GtK*{1830}            \\ \hline
		\PKOne[1270] & \MtKOne*{1270}    & \GtKOne*{1270}         \\
		\PKOnePr     & \MtKOne*{1650}    & \GtKOne*{1650}         \\ \hline
		\PKTwo[1770] & \MtKTwo*{1770}    & \GtKTwo*{1770}         \\
		\PKTwo[1820] & \MtKTwo*{1820}    & \GtKTwo*{1820}         \\
		\PKTwo[2250] & \MtKTwo*{2250}    & \GtKTwo*{2250}         \\ \hline
		\PKTwoSt     & \MtKTwoSt*{1430}  & \GtKTwoSt*{1430}       \\ \hline
		\PKThree*    & \MtKThree*{2320}  & \GtKThree*{2320}       \\ \hline
		\PKFour*     & \MtKFour*{2500}   & \GtKFour*{2500}        \\ \hline
		\PKFourSt    & \MtKFourSt*{2045} & \GtKFourSt*{2045}
	\end{tabular}

\end{table}%

\newcommand\wavescaption{Top row: intensity spectra of selected partial waves. Bottom row: real part $\Re(\sdm*)$ of the \WaveK4+1+\Prho\PK G wave (d), and phase of the \WaveK0-0+\Prho\PK P wave (e), both relative to the \WaveK1+0+\Prho\PK S wave. Data points represent the measurements. Curves represent the \textcolor{MpRed}{total \rmfmodel}~(\textcolor{MpRed}{red}), the individual \textcolor{MpBlue}{resonance components}~(\textcolor{MpBlue}{blue}), the \textcolor{MpGreen}{non-resonant components}~(\textcolor{MpGreen}{green}), the \textcolor{MpOrange}{\threePi background components}~(\textcolor{MpOrange}{orange}), and the \textcolor{MpBrown}{\ebkg components}~(\textcolor{MpBrown}{brown}). Extrapolations of the model curves beyond fitted \mKpipi ranges are shown in lighter colors, grey for data points. The inserts in (a) and (c) show the intensity spectrum in log scale. All figures show the range of \SIvalRange{0.1}{\tpr}{0.15}{\GeVcsq}.}
\begin{figure*}[tb]%
	\centering%
	\hfill%
	\subfloat[]{\includegraphics[width=0.33\linewidth]{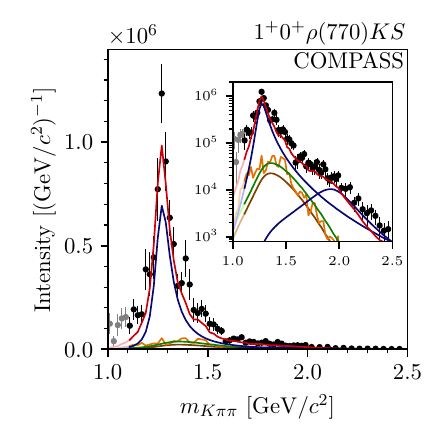}\label{fig:waves:1p}}%
	\subfloat[]{\includegraphics[width=0.33\linewidth]{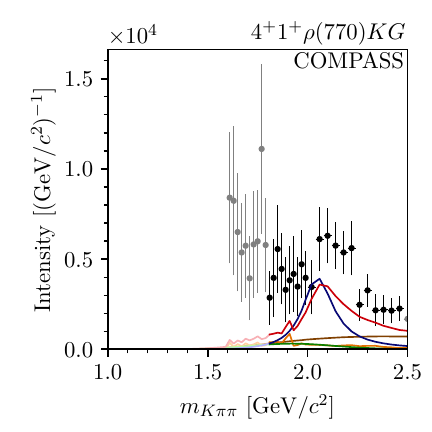}\label{fig:waves:4p}}%
	\subfloat[]{\includegraphics[width=0.33\linewidth]{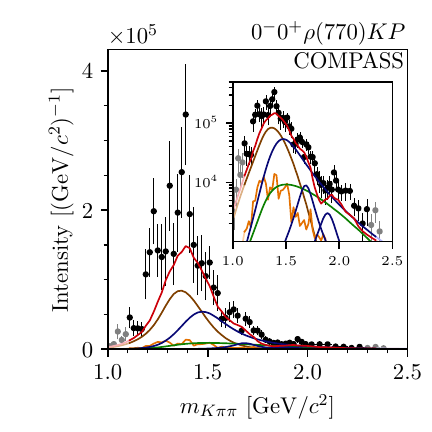}\label{fig:waves:0mi}}%
	\\\hfill%
	\subfloat[]{\includegraphics[width=0.33\linewidth]{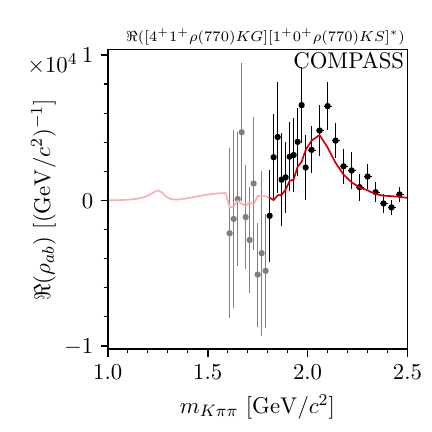}\label{fig:waves:4pre}}%
	\subfloat[]{\includegraphics[width=0.33\linewidth]{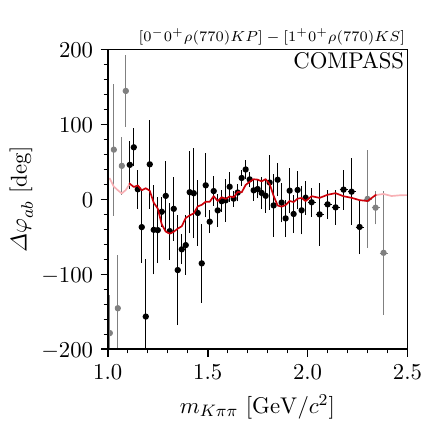}\label{fig:waves:0mp}}%
	\ifdef{\onecolumn}{}{%
		\begin{minipage}{0.33\linewidth}%
			\vspace*{-18em}%
			\caption{\wavescaption}%
		\end{minipage}%	
	}
	\ifdef{\onecolumn}{\caption{\wavescaption}}{}
	
	%\caption{Intensity spectra of selected partial waves (a)--(c) and phase of the \WaveK0-0+\Prho\PK P wave relative to the \WaveK1+0+\Prho\PK S wave (d) in the first \tpr bin. The black data points represent the measured intensities. The \textcolor{MpRed}{red} curves represent the \textcolor{MpRed}{total \rmfmodel}. The \textcolor{MpBlue}{blue} curves represent the individual \textcolor{MpBlue}{resonance components}, the \textcolor{MpGreen}{green} curves the \textcolor{MpGreen}{non-resonant components}, the \textcolor{MpOrange}{orange} curves the \textcolor{MpOrange}{\threePi background components}, and the \textcolor{MpBrown}{brown} curves the \textcolor{MpBrown}{\ebkg components}. The extrapolations beyond the \mKpipi fit ranges are shown in lighter colors. The corresponding data points are shown in gray.}%
	\label{fig:results}%
\end{figure*}

Overall, the resonance-model fit reproduces well the spin-density matrix elements of the 14 selected waves as illustrated exemplarily in \cref{fig:waves:1p}. The result for all spin-density matrix elements can be found in Figs. 3--9 of \refCite{supplemental}.\todoGr{Check supplemental figure number}
\Cref{tab:parameters} lists the parameters of the eleven measured resonances, which we discuss below, starting with established states.

%=============================================================================
%=============================================================================
\subsection{Established \texorpdfstring{\PKOne*}{K1} and \texorpdfstring{\PKTwoSt*}{K2*} States}

%We now discuss \PKOne* states in \WaveK1+M+\Prho\PK S  with spin projections $M=0, 1$. Both exhibit a clear \PKOne[1270] signal.
%We determined its mass and width to be \MKOne{1270} and \GKOne{1270}. In neither wave we observe any significant signal of the \PKOne[1400], consistent with previous observations~\cite{ParticleDataGroup:2024cfk}\todoBl{Stephan: {\color{red} expect that we would observe or that we don't observe ? So we insert this by hand ? does is exhibit any strength ? we should say something to this}\\Stefan: I dropped the sentence that we fix the K1(1400) parameters}.
%In the high-mass tail of the \PKOne[1270], the intensity exhibits a shoulder, which we modeled by a \PKOnePr component, and measured its mass and width to be \MKOne{1650} and \GKOne{1650} respectively.
%

Our \PKOne[1270] width agrees well with the PDG average.
Although our \PKOne[1270] mass is slightly larger than the PDG average it is still consistent with it and also with several previous measurements~\cite{ACCMOR:1981yww,Gavillet1978,LHCb:2017swu,Rodeback1981} and with a phenomenological model assuming a two-pole structure of the \PKOne[1270]~\cite{Geng2007}.
% In addition, the authors of \refCite{Geng2007} suggested that the \PKOne[1270] has a two-pole structure, with the heavier pole coupling predominantly to the $\Prho\PK$ decay channel, in agreement with our measurement.
%
We do not observe a significant \PKOne[1400] signal in the selected $\JP=\oneP$ waves. This is expected, since the ambiguous identification discussed above limits the analysis to the $\Prho\PK$ decay and the \PKOne[1400] has a small branching fraction of this decay~\cite{ParticleDataGroup:2024cfk}, which is at our detection limits.

In the mass range from 1.7 to \SI{1.9}{\GeVcc}, quark models predict two excited \PKOne* states. However, the stability of the \rmf allowed only a single excited \PKOne* Breit-Wigner component dubbed \PKOnePr.
LHCb reported two excited \PKOne* in the decay $B^+ \to \Jpsi\Pphi*\PKPos$~\cite{LHCb:2021uow}. The parameters of the heavier of these two states agree well with our \PKOnePr signal.
The PDG lists only a single excited \PKOne* called \PKOne[1650].
The PDG mass estimate is based on only a single measurement~\cite{Frame:1985ka} and is much smaller than the mass of our signal and of the two states observed by LHCb.
%However, in \refCite{Frame:1985ka} they only considered the intensity of the \oneP wave, ignoring phase information.

%This supports the existence of an excited \PKOne* state around \SI{1.9}{\GeVcc}.\todo{does it support our measurement or the LHCb measurement ?}

%We discuss \PKTwoSt* states in the \WaveK 2+1+\PKSt\Ppi D and \WaveK 2+1+\Prho\PK D waves. Both show a clear \PKTwoSt. We measured its mass to \MKTwoSt{1430} and its width to \GKTwoSt{1430}. In neither wave we observe  evidence for an excited \PKTwoSt*.
%
% Masses and width for the \PKTwoSt are listed individually for both charges states~\cite{ParticleDataGroup:2024cfk}.
% %The PDG lists the average mass and width values separately for the charged and neutral \PKTwoSt.
% Our result for the width of the charged \PKTwoSt agrees well with the average of the charged state, however we obtain a mass slightly larger than average. In turn, it agrees well with the average value of the neutral state and with individual measurements~\cite{Ablikim2019,Cleland1982,Martin1978} of the charged state. This may indicate that the separation into averages for the charged and neutral \PKTwoSt by the PDG is artificial.{\color{red}~I am not sure what we want to say here...}
Our \PKTwoSt mass and width agree well with the PDG averages. The corresponding uncertainties are competitive with previous measurements of the \PKTwoSt.
%while again our mass is slightly larger, but in good agreement with various individual measurements~\cite{ParticleDataGroup:2024cfk}.{\todo{sounds funny.. we always we agree well and than we make th point that our mass is higher...}} %~\cite{Ablikim2019,Cleland1982,Martin1978,Barnham:1971jc,Schweingruber:1968zz,Bossano:1967yr}.

%=============================================================================
%=============================================================================
\subsection{Ground and Excited \texorpdfstring{\PKTwo*}{K2} States}

%We studied \PKTwo* resonances by simultaneously fitting four \twoM waves in the \rmf; \ie \WaveK2-0+\PKTwoSt\Ppi S, \WaveK2-0+\PfTwo\PK S, \WaveK2-0+\PKSt\Ppi F, and \WaveK2-0+\Prho\PK F.
%All four waves exhibit a broad intensity peak at about \SI{1.7}{\GeVcc} \see{fig:waves:2mi} accompanied by a rising phase. The exact peak position and shape is different in the four waves and changes as a function of \tpr indicating that the peak arises from the interference of multiple components.
%%
%The \SI{1.7}{\GeVcc} region is reproduced well by the \rmf by including a \PKTwo[1770] and a \PKTwo[1820] component.
%We measured mass and width for the \PKTwo[1770] to be \MKTwo{1770} and \GKTwo{1770}, respectively, and \MKTwo{1820} and \GKTwo{1820}, correspondingly for the \PKTwo[1820].

Quark models predict two \PKTwo* ground states in the \SI{1.7}{\GeVcc} mass region.
However, most previous experiments have identified only a single state.
Only our measurement and previous measurements by LHCb~\cite{LHCb:2016axx} and LASS~\cite{Aston1993} could resolve two states, \PKTwo[1770] and \PKTwo[1820]. LHCb data favor the 2-state hypothesis with a significance of 5.8~ $\sigma$. Our measurement yields an even higher statistical significance of 11~$\sigma$, estimated from the $\chi^2$ difference \wrt the single-resonance hypothesis using Wilks' theorem~\cite{Wilks:1938dza}. This clearly excludes the single-resonance hypothesis.
Our resonance parameters agree well with both previous measurements, except for the \PKTwo[1770] mass, which is consistent with LHCb, but significantly lower than the LASS result. However, it is not clear whether the uncertainties of the LASS measurement include systematic effects.
%
%The intensities of the \WaveK2-0+\PKTwoSt\Ppi S and \WaveK2-0+\PfTwo\PK S waves exhibit a distinct high-mass shoulder above \SI{2}{\GeVcc} accompanied by a rise in the relative phase of these waves. Both are reproduced well in the \rmf by a \PKTwo[2250] component. We measured mass and width to be \MKTwo{2250} and \GKTwo{2250}, respectively.
%
For the excited \PKTwo*, our \PKTwo[2250] mass and width agree with the corresponding PDG average values, which are based only on analyses of  $\Lambda \widebar p$ / $\widebar\Lambda p$ final states.
Our simultaneous measurement of three \PKTwo* resonances with masses below and above \SI{2}{\GeVcc} from the simultaneous fit to four partial waves representing the $\PKTwoSt\Ppi$, $\PfTwo\PK$, $\PKSt\Ppi$, and $\Prho\PK$ decays is the most comprehensive single analysis of the strange $\JP=\twoM$ sector.

%=============================================================================
%=============================================================================
\subsection{\texorpdfstring{\PKThree*}{K3} and \texorpdfstring{\PKFour*}{K4} States}

%We discuss \PKThree* states in the \WaveK 3+0+\PKThreeSt\Ppi S and \WaveK3+1+\PKTwoSt\Ppi P waves. Both exhibit an intensity peak at about \SI{2.1}{\GeVcc}, accompanied by a rise of their relative phase.
%We included a \PKThree* component and measured its mass to be \MKThree{2320} and its width to be \GKThree{2320}.
%%
%We discuss \PKFour* states in the \WaveK4-0+\PKTwoSt\Ppi D wave, which exhibits a broad bump in the intensity distribution at about \SI{2.3}{\GeVcc}, accompanied by a rise of its relative phase.
%We included a \PKFour* resonance in our model and measured its mass to be \MKFour{2500} and its width to be \GKFour{2500}.
%%
%Our \rmf describes all the off-diagonal spin density matrix elements for the \threeP and \fourM waves well. Similar to the \fourP waves, we underestimate their intensities, 
%as their contribution to the total intensity again is only in the per mil range.

We measure the masses of a \PKThree* and a \PKFour* state, both in good agreement with quark-model predictions for the \PKThree* and \PKFour* ground states~\cite{Ebert2009,Oudichhya:2023lva,Taboada-Nieto:2022igy,Pang:2017dlw,Godfrey1985} (see \cref{fig:kaonSpectrum}).
The only \PKThree* and \PKFour* states listed by the PDG are \PKThree and \PKFour. Both have only been observed in the $\Lambda \widebar p$ / $\widebar\Lambda p$ decay, the \PKThree by the CERN $\Omega$ spectrometer~\cite{Bari-Birmingham-CERN-Milan-Paris-Pavia:1983nwf} and the Geneva-Lausanne spectrometer~\cite{Cleland:1980ya}, the \PKFour by the latter one only~\cite{Cleland:1980ya}.
However, the PDG masses for the \PKThree and \PKFour are about \SI{200}{\MeVcc} larger than the masses of our \PKThree* and \PKFour* signals, and match better with quark-model predictions for the corresponding first excited states.
This suggests that we have identified the \PKThree* and \PKFour* ground states, while the previous measurements may have observed the corresponding first excitations.

%=============================================================================
%=============================================================================
\subsection{Established \texorpdfstring{\PKFourSt}{K4*(2045} State}
%We discuss \PKFourSt* states in the \WaveK4+1+\PKSt\Ppi G and \WaveK4+1+\Prho\PK G waves. Both exhibit an intensity peak and a rising phase around \SI{2}{\GeVcc}, which we modeled by a \PKFourSt component.
%

Our measurements of the \PKFourSt mass and width are the most accurate and agree with the PDG averages. 
The corresponding off-diagonal elements of $\rho_{ab}$ are well reproduced (\cref{fig:waves:4pre}).
Our \rmf reproduces the shape of the intensity given by $\rho_{aa}$ fairly well, but underestimates it, as shown in \cref{fig:waves:4p}.
This is observed for several waves with relative intensities at the per mil level.
The off-diagonal elements of $\rho_{ab}$ representing interference with large waves are well constrained by data and encode information about the magnitude and phase of the partial wave.
The measured intensities of small waves, however, are more weakly constrained as they contribute less to the modeled distribution of events in \cref{eq:intensity}.
As a result, the intensities are susceptible to analysis artifacts, \eg caused by the ambiguous identification mentioned above.
%For some waves that only contribute to the total intensity at the per mil level, such as for the $\JP=\fourP$ waves, the \rmf approximates the mass dependence of the wave intensity, but underestimates its total intensity (see \cref{fig:waves:4p}). In contrast, the off-diagonal elements of $\rho_{ab}$, which encode the full information about the partial wave amplitudes, are well reproduced (see \cref{fig:waves:4pre}).
%
%This imperfection in the modeling of intensities is due to the low constraints on the measured intensities of small waves in \cref{eq:intensity}, while their off-diagonal $\rho_{ab}$ elements are more constrained by interference from large waves. 

%
The \rmf is mainly driven by the off-diagonal $\rho_{ab}$ elements, which are fit well. We therefore conclude that any potential bias in the resonance parameters from underestimating the intensities is small and covered by the quoted systematic uncertainties.
The accurate measurement of the well-known \PKFourSt and its robustness in systematic studies demonstrates our sensitivity to signals at the per mil level, despite the imperfections in modeling their intensity, and thus validates our analysis results.

%=============================================================================
%=============================================================================
\subsection{First Observation of a Crypto-Exotic \texorpdfstring{$\JP=\zeroM$}{J\textsuperscript{P}=0\textsuperscript{-}} State}

\settw 0-0+\Prho\PK P
Excited pseudoscalar strange mesons are best seen in our \tw wave shown in \cref{fig:waves:0mi}. We observe three resonance structures with masses around \SI{1.4}{\GeVcc}, \SI{1.7}{\GeVcc}, and \SI{1.9}{\GeVcc}.
%with the lowest peak at about \SI{1.4}{\GeVcc}, a second peak at about \SI{1.7}{\GeVcc} and a small bump at about \SI{1.9}{\GeVcc}.
%We modeled this wave by three resonance components.
%one corresponding to the established \PK[1460], one to the elusive \PK[1630], and one to the unestablished \PK[1830].
The low-mass region of $\mKpipi\lesssim\SI{1.5}{\GeVcc}$ in this wave is weakly affected by the ambiguous identification discussed above.
We therefore fixed the resonance parameters of the \PK[1460] to the values obtained by LHCb~\cite{LHCb:2017swu}, the measurement with smallest uncertainties.
%We cannot reliably measure its amplitude in this mass region and have therefore fixed the resonance parameters of the \PK[1460] component to the values from LHCb~\cite{LHCb:2017swu}.

Above the \PK[1460], the pronounced peak at \SI{1.7}{\GeVcc} is accompanied by a rise of the relative phase of the \tw wave, exemplarily shown in \cref{fig:waves:0mp}. The intensities and phases are well reproduced by a resonance component dubbed \PK[1690]. Again, we attribute imperfections in modeling the intensity to effects discussed above for the $\JP=\fourP$ waves.
We estimated the statistical significance of the \PK[1690] to be about $8\sigma$.%{\todo{Stephan: lets see the real part.. if this is described well, we can reformulate this sentence//Stefan: The real part can be seen in the spin-density matrices in the appendix/supplemental}}.
%
%The PDG lists a \PK[1630] state from only a
%A \PK[1630] state has only been observed in a single measurement using the reaction $\PpiNeg p \to (K^0\sub{S} \PpiPos \PpiNeg) X^+\PpiNeg X^0$~\cite{Karnaukhov:1998qq,Karnaukhov2000}.\todo{Or completely disregard this measurement - lets have a look at the quality of the data}
%The authors did not perform a partial-wave analysis, but only identified peaks in the mass spectrum. They were unable to determine the quantum numbers of the signal~\cite{Karnaukhov2000}. In addition, they estimated a width of only about \SIaerr{16}{19}{-16}{\GeVcc}, which is considerably smaller than expected and typically observed for a strange-meson resonance. This casts doubts on the validity and interpretability of the observation from \refsCite{Karnaukhov:1998qq,Karnaukhov2000}.
%Although our mass for the \PK[1630] agrees with \refsCite{Karnaukhov:1998qq,Karnaukhov2000}, our measured width is larger and more typical for a strange-meson resonance.

A \PK[1630] has been claimed by \refsCite{Karnaukhov:1998qq,Karnaukhov2000} using bubble-chamber data on the reaction $\PpiNeg p \to (K^0\sub{S} \PpiPos \PpiNeg) X^+\PpiNeg X^0$ without performing a partial-wave analysis. Consequently, the authors did not determine the quantum numbers of their signal.
To extract a peak, kinematic cuts had been applied restricting the phase space of the selected events without further justification.
The reported width of \SIaerr{16}{19}{-16}{\MeVcc} is unusually small.
%\todoBl{Stefan: "and probably in the range of their mass resolution (not quoted by the authors): Should we speculate about their mass resolution (while I see your point, which additionally weakens this measurement)?}
Thus we do not consider their observation in the context of our \PK[1690].
%\todoBl{Stefan: Should we remove the K(1630) as well from Fig1? Stephan: Lets discus the quality of their measurements among ourselves in detail. Then we can also mention that we omitted this state form the figure.}

%\todoInl{Stephan: the ACCMOR discussion should not take our space. it is not our task to reveal a signal in other peoples data or speculate about it. }
%In addition, one may find indications for a \PK[1630] signal in the data from \refCite{ACCMOR:1981yww}. Their measurement of the intensity and relative phase of the \WaveK 0-0+\Prho\PK P wave is shown in figure 18 in \refCite{ACCMOR:1981yww}. Their model includes a \PK[1460] only and the resulting fit badly describes the data at about \SI{1.7}{\GeVcc}. This leaves room for a possible signal from \PK[1630]. However, this issue is not addressed by the authors of \refCite{ACCMOR:1981yww}.

The RMF also yields the \PK[1830], which corresponds to the small intensity bump at \SI{1.9}{\GeVcc}.
The statistical significance is about $5\sigma$. LHCb~\cite{LHCb:2016axx} has reported evidence for a similar signal, but our resonance parameters are more than twice as precise.\todoBl{The LHCb measurement has $3.5\sigma$ significance}
%The mass and width of the \PK[1830] have only recently been measured reliably by LHCb~\cite{LHCb:2016nsl,LHCb:2016axx}. Their result agrees well with our much more precise measurement.

In total, our \tw wave exhibits evidence for three resonances signals.\todoBl{Stefan: The K(1690) has about $8\sigma$, but the K(1830) has only about $5\sigma$. Due to the limited applicability of Wilk's theorem here, we cannot state that the significance for the K(1830) is above $5\sigma$, only above $3\sigma$. This is all given in the paragraphs above. Here, I would say that we have "only" evidence for three states, consequently.}
We identified these three signals in a single self-consistent analysis, fixing the lightest of the states to the values reported by LHCb. The significance of our observations and the measured resonance parameters are robust to all systematic studies. The presence of three states does not match quark-model predictions of only two states in this mass region~\cite{Ebert2009,Oudichhya:2023lva,Taboada-Nieto:2022igy,Pang:2017dlw,Godfrey1985}.
This indicates the presence of a supernumerary signal.
The lighter of the two quark-model states is consistent with the \PK[1460], the heavier matches best the \PK[1830]. This leaves our observation of a \PK[1690] as the first clear candidate for a cyrpto-exotic strange meson with $\JP=\zeroM$.
One should note that the observed states may be mixtures of pure quark-model and exotic states, and that a resonance-like structure, \ie an intensity peak accompanied by a rising phase, can also be caused by kinematic singularities, including triangle singularities~\cite{Guo2020,Alexeev2021}.
%This may be tested for in the future, e.g. using $\tau$ lepton decays \cite{Rabusov:2024koz}.\todo{Stefan: Do we need this a1 disucssion here, except as ad for the tau analysis.}

%\begin{figure}%
%	\centering%
%	\includegraphics{plots/rmf/main/rho/rho__0-__tPrime_0.100_0.150.pdf}%
%	\caption[Extended \rmf results for the spin-density matrix of the \WaveK0-0+\Prho\PK P wave.]{Same as \cref{fig:results:1p:sdm}, but showing the results of an extended \rmf, where the \WaveK0-0+\Prho\PK P wave was included in addition to the waves of the 10-wave \rmf. The \WaveK 1+0+\Prho\PK S and \WaveK2+1+\Prho\PK D waves serve as phase references. The second-lowest \tpr bin is shown.}%
%	\label{fig:results:0m:sdm}%
%\end{figure}

%=============================================================================
%=============================================================================
%=============================================================================
\section{Conclusions}

%based on the world's largest data sample of the reaction \reactionK, 
%We have performed the most detailed and self consistent partial wave analysis of the \Kpipi system. The measurement of masses and widths of 11 strange-meson resonances covering the mass region of \SIrange{1.2}{2.4}{\GeVcc} resulted in the most complete picture for the strange meson spectrum. We notably improve the uncertainty for some of these parameters, in particular for mass and width of the \PK[1830].
Our detailed and self-consistent partial-wave analysis of the world's largest data sample of the reaction \reactionK yields the masses and widths of eleven strange-meson resonances in the mass region from \SIrange{1.2}{2.4}{\GeVcc}. This is the most complete measurement of the strange-meson spectrum from a single analysis.
We improve the resonance parameters for many states, in particular for the \PK[1830].
%
%In summary, we performed the most comprehensive and detailed partial-wave analysis of the \Kpipi system to date. This resulted in the most complete picture of the strange-meson spectrum backed by a self-consistent analysis. We measured the masses and widths of 11 strange-meson resonances covering the mass region of \SIrange{1.2}{2.4}{\GeVcc}. We notably improve the uncertainty for some of these parameters, in particular for mass and width of the \PK[1830].
%
Our results also suggest that we have uncovered the \PKThree* and \PKFour* ground states for the first time.
%We observe a \PKThree* signal and have evidence for a \PKFour* signal, which are both consistent with predictions for the corresponding ground states for the first time.
%
Most notably, we have identified the \PK[1690], a supernumerary pseudoscalar resonance signal with a mass of about \SI{1.7}{\GeVcc}. This signal lies in mass between the two predicted quark-model states, which we also observe, and is the first candidate for a crypto-exotic strange meson with $\JP=\zeroM$.
%A highlight from this analysis are indications for an exotic strange meson in the pseudoscalar sector, manifesting itself as a supernumerary resonance 

We gratefully acknowledge the support of the CERN management and staff and the skill and effort the technicians of our collaborating institutes.
This work was made possible by the financial support of our funding agencies.
This work was supported by the Excellence Cluster Universe which is funded by the Deutsche Forschungsgemeinschaft (DFG, German Research Foundation), the Excellence Cluster ORIGINS which is funded by the DFG under Germany's Excellence Strategy -- EXC 2094 -- 390783311, the Computational Center for Particle and Astrophysics (C2PAP), and the Leibniz Supercomputer Center (LRZ).

% Stefan Wallner, Godo Kurten: Supported by the Max Planck Institute for Physics, 85748 Garching, Germany

%We are indebted to the Max Planck Institute for Physics for its support during the final analysis phase.

% Present address: Max Planck Institute for Physics, 85748 Garching, German

% This work was supported in part by the Deutsche Forschungsgemeinschaft (DFG, German Research Foundation) through the Excellence Clusters UNIVERSE and ORIGINS under Germany's Excellence Strategies EXC-2094-390783311.

\clearpage
%\appendix
\addcontentsline{toc}{section}{Supplemental Material}
%=============================================================================
%=============================================================================
%=============================================================================

\supplementalsection{Construction of the Resonance Model}

The first stage of the analysis, called partial-wave decomposition, is described in detail in \refCite{Wallner:2022prx}. The results of the partial-wave decomposition presented in that reference are identical to those presented in this paper. The second analysis stage, \ie the resonance-model fit (RMF), presented in this paper deviates from what is presented \refCite{Wallner:2022prx} in terms of the selected set of 14 partial waves and the construction of the resonance model, while the method is the same. Therefore, we briefly describe the construction of the resonance model used in this paper in the following.

The 14 selected partial waves included in the RMF are parameterized by in total 13 resonance components, as well as a non-resonant and two background components for each wave. \Cref{tab:appendix:rmf:waves} lists these waves and the included components. In the following, we briefly summarize the parameterizations used for the individual model components. Details can be found in \refCite{Wallner:2022prx}.

\newcommand{\tabResonanceList}[2]{\multirow#1{3.52cm}{\centering$\cBrk[3]{\mkern-10mu\begin{array}{c}#2\end{array}\mkern-10mu}$}}
\begin{table}[!b]%
	\setTableStretchDef%
	\centering%
	\captionLong{List of partial waves and model components included in the \rmf.}{The second column lists the included resonance components. Their parameters are specified in \cref{tab:appendix:rmf:res}. The next three columns list the parameterization for the dynamic amplitudes of the non-resonant components (NR), the model used for the \threePi background components and the parameterizations used for the dynamic amplitudes of the \ebkg components (\ebkg*). The numbers refer to the equation numbers in the text. The last two columns list the mass range for a partial wave included in the \rmf.}%
	\label{tab:appendix:rmf:waves}%
	\begin{tabular}{l|clcl|cc}
		\toprule%
		\multirow2*{Partial Wave} &                           \multirow2*{Resonances}                           &            \multirow2*{NR}            & \multirow2*{\threePi}  &           \multirow2*{\ebkg*}           & \multicolumn{2}{c}{\mKpipi Range}    \\
		& & & & & \multicolumn{2}{c}{\si\GeVcc}\\ \midrule
		\WaveK 0-0+\Prho\PK P        & \shortstack{\PK[1460], \PK[1690],\\\PK[1830]}                               & \eqref{eq:rmf:model:nrs} & \eqref{eq:rmf:model:3pi:sdm} & \eqref{eq:rmf:model:bkgs} & 1.10 &            2.30              \\\hline
		\WaveK 1+0+\Prho\PK S        & \tabResonanceList{2}{\PKOne[1270], \PKOne[1400],\\\PKOnePr} & \eqref{eq:rmf:model:nrs} & \eqref{eq:rmf:model:3pi:sdm} & \eqref{eq:rmf:model:bkgs} & 1.10 &            2.50              \\
		\WaveK 1+1+\Prho\PK S        &                                                                & \eqref{eq:rmf:model:nrs} & \eqref{eq:rmf:model:3pi:sdm} & \eqref{eq:rmf:model:bkg} & 1.10 &            2.50              \\\hline
		\WaveK 2+1+\PKSt\Ppi D       &                 \tabResonanceList{2}{\PKTwoSt}                 & \eqref{eq:rmf:model:nrs} & \eqref{eq:rmf:model:3pi:sdm} & \eqref{eq:rmf:model:bkgs} & 1.20 & 1.70      \\ 
		\WaveK 2+1+\Prho\PK D       &                                  & \eqref{eq:rmf:model:nrs} & \eqref{eq:rmf:model:3pi:sdm} & \eqref{eq:rmf:model:bkgs} & 1.30 & 1.70      \\ \hline
		\WaveK 2-0+\PKSt\Ppi F & \tabResonanceList{4}{\PKTwo[1770], \PKTwo[1820],\\\PKTwo[2250]}  & \eqref{eq:rmf:model:nrs} & \eqref{eq:rmf:model:3pi:sdm} & \eqref{eq:rmf:model:bkgs} & 1.60 & 2.00 \\ 
		\WaveK 2-0+\PKTwoSt\Ppi S &   & \eqref{eq:rmf:model:nr} & \eqref{eq:rmf:model:3pi:sdm} & \eqref{eq:rmf:model:bkg} & 1.50 & 2.80 \\
		\WaveK 2-0+\Prho\PK F &   & \eqref{eq:rmf:model:nrs} & \eqref{eq:rmf:model:3pi:sdm} & \eqref{eq:rmf:model:bkgs} & 1.60 & 2.10 \\
		\WaveK 2-0+\PfTwo\PK S &   & \eqref{eq:rmf:model:nrs} & \eqref{eq:rmf:model:3pi:sdm} & \eqref{eq:rmf:model:bkgs} & 1.60 & 2.80 \\\hline
		\WaveK 3+0+\PKThreeSt\Ppi S  & \tabResonanceList{2}{\PKThree*} & \eqref{eq:rmf:model:nrs} & \eqref{eq:rmf:model:3pi:sdm} & --- & 2.00 &            2.50              \\
		\WaveK 3+1+\PKTwoSt\Ppi P  &  & \eqref{eq:rmf:model:nrs} & \eqref{eq:rmf:model:3pi:sdm} & \eqref{eq:rmf:model:bkgs} & 2.00 &            2.50              \\\hline
		\WaveK 4+1+\PKSt\Ppi G & \tabResonanceList{2}{\PKFourSt} & \eqref{eq:rmf:model:nrs} & \eqref{eq:rmf:model:3pi:sdm} & \eqref{eq:rmf:model:bkgNoPS} & 1.80 & 2.50 \\
		\WaveK 4+1+\Prho\PK G &  & \eqref{eq:rmf:model:nrs} & \eqref{eq:rmf:model:3pi:sdm} & \eqref{eq:rmf:model:bkgs} & 1.80 & 2.50 \\\hline
		\WaveK 4-0+\PKTwoSt\Ppi D & \PKFour* & \eqref{eq:rmf:model:nrs} & \eqref{eq:rmf:model:3pi:sdm} & \eqref{eq:rmf:model:bkgs} & 2.10 & 2.80 \\
		\bottomrule
		% &  &  &  &  &
	\end{tabular}%
\end{table}

%\todoInl{Change non-resonant parameter labels away from $a$}

For the parameterization of the dynamic amplitudes of resonance components relativistic Breit-Wigner amplitudes are used~\cite{PhysRev.49.519,Perl:1974}
\begin{equation}
	\isobarDynamicF[\mathrm{BW}][][m][; m_0, \Gamma_0] = \frac{m_0\Gamma_0}{m_0^2 - m^2 - i\,m_0\, \Gamma(m)},\label{eq:pwd:model:bw}
\end{equation}
with the mass-dependent width
\begin{equation}
	\Gamma(m) = \Gamma_0 \frac{q_i(m)}{m} \frac{m_0}{q_i(m_0)} \frac{\barrierFactor[L_i][2]{m}}{\barrierFactor[L_i][2]{m_0}},\label{eq:pwd:model:mdepwidth}
\end{equation}
which takes into account the opening of the phase space for the decay mode $i$ in the two-body approximation.
The two-body break-up momentum,
\begin{equation}
	q_i(m) = q(m, m_1, m_2) = \frac{\sqrt{\sBrk{m^2-(m_1+m_2)^2}\sBrk{m^2-(m_1-m_2)^2}}}{2 m}, \label{eq:pwd:model:breakupmomentum}
\end{equation}
is given by the masses $m_1$ and $m_2$ of the daughter particles.
\barrierFactor[L_i]{m} is the centrifugal barrier factor, where $L_i$ is the orbital angular momentum between the daughter particles.\footnote{We used the parameterization from von Hippel and Quigg~\cite{VonHippel:1972fg}. See appendix D in \refCite{Ketzer:2019wmd} for the definition of the centrifugal-barrier factors as a function of $z=q^2(\mKpipi, \mhh, m_{\bachelor})/q^2\sub{R}$, where $q$ is the two-body break-up momentum of the $\X \to \isobar \bachelor$ decay \seeEq{eq:pwd:model:breakupmomentum} and $q\sub{R}=\SI{197.3}{\MeVc}$.}
\Cref{tab:appendix:rmf:res} lists the decay modes used for the dynamic widths, the parameter limits applied in the fit, and the start parameter ranges that are used when randomly generating start parameters for the RMF.

The employed parameterization for the non-resonant components,
\begin{equation}
	\dynamicFNR = (\mKpipi - m\sub{thr})^{\shapeParameterNRa} e^{-b(\shapeParameterNRc)\, \effTwoBodyBM[\mKpipi][2]}, \label{eq:rmf:model:nr}
\end{equation}
is inspired by \refCite{Tornqvist1995} and was used in previous analyses, such as the COMPASS \threePi analysis~\cite{COMPASS:2018uzl}.

Here, $m\sub{thr} = m_K + 2m_{\pi}$ is the kinematic threshold for \mKpipi, and\footnote{We used $m\sub{norm} = \SI{3}{\GeVc}$.}
\begin{equation}
	\effTwoBodyBM[\mKpipi][] = q(m\sub{norm}, m_{\isobar*}, m_{\bachelor*}) \frac{\mKpipi \decayNormIntegral[\waveLabelA(\componentLabelK)]}{m\sub{norm} \decayNormIntegral[\waveLabelA(\componentLabelK)][][m\sub{norm}]}, \label{eq:rmf:model:twobodybreakupapprox}
\end{equation}
is an extension of the two-body break-up momentum of the isobar-bachelor system, which takes into account the finite width of the isobar via the phase-space integral \decayNormIntegral[\waveLabelA(\componentLabelK)] of wave \waveLabelA in which component \componentLabelK is modeled, and is hence valid also below the nominal two-body threshold (see section~6.1.2 in \refCite{Wallner:2022prx} for details).
The free shape parameters are \shapeParameterNRa and \shapeParameterNRc, where we required the slope parameter $b(\shapeParameterNRc)$ of the exponential in \cref{eq:rmf:model:nr} to be larger than $\SI{-1}{\perGeVcsq}$ by using the following parameter mapping:
\begin{equation}
	b(\shapeParameterNRc) = \sbrk{-1 + e^{\shapeParameterNRc}} \si\perGeVcsq.% \SI{-1}{\perGeVcsq} + \SI{1}{\perGeVcsq} \cdot e^{\shapeParameterNRc}.
\end{equation}
For most of the studied partial waves, a simplified version of \cref{eq:rmf:model:nr} with $\shapeParameterNRa = 0$ and $b(\shapeParameterNRc) = \shapeParameterNRb$, \ie
\begin{equation}
	\dynamicFNRs = e^{-\shapeParameterNRb\,\effTwoBodyBM[\mKpipi][2]}, \label{eq:rmf:model:nrs}
\end{equation}
turned out to be sufficient to describe the non-resonant components.

The \threePi background component is parameterized by
\begin{equation}
	\sdm[\waveLabelA][\waveLabelB][3\pi] = \Abs[2]{\coupling*[][\threePiShort][][]}^2 \tilde{\rho}^{3\pi}_{ab}\mKpipiTpr. \label{eq:rmf:model:3pi:sdm}
\end{equation}
Here, $\tilde{\rho}^{3\pi}_{ab}\mKpipiTpr$ is the spin-density matrix of \Kpipi partial waves obtained from a fit to simulated \threePi data generated according to the result of the partial-wave decomposition of the COMPASS \threePi data~\cite{COMPASS:2015gxz}.
The absolute amount of \threePi background in our sample is given by the free parameter $\abs{\coupling*[][\threePiShort][][]}^2$. This means $\abs{\coupling*[][\threePiShort][][]}^2$ is determined from the measured \Kpipi sample, while \sdm[\waveLabelA][\waveLabelB][\threePiShort] is completely determined by the measured \threePi sample.

Other incoherent background processes, such as \reactionKKK, also contribute to the \Kpipi sample.
As there are no explicit models available for these processes, we parameterized them in an effective way by using the same phenomenological functional dependence on \mKpipi as used for the non-resonant components.
The \rmf model for the \ebkg components reads
\begin{equation}
	\rmfTransitionAmpl[\waveLabelA][\ebkg*] =  \mathcal{K}\mKpipiTpr \,\coupling[\waveLabelA][\ebkg*][] \, \dynamicFBkg[\componentLabelK_\waveLabelA]. \label{eq:rmf:model:ebkg:trans}
\end{equation}
For most of the partial waves, we used the simplified parameterization [same as \cref{eq:rmf:model:nrs}]:
\begin{equation}
	\dynamicFBkgs = e^{-\shapeParameterNRb\,\effTwoBodyBM[\mKpipi][2]}. \label{eq:rmf:model:bkgs}
\end{equation}
For some partial waves we used the full parameterization [same as \cref{eq:rmf:model:nr}]:
\begin{equation}
	\dynamicFBkg = (\mKpipi - m\sub{thr})^{\shapeParameterNRa} e^{-b(\shapeParameterNRc)\, \effTwoBodyBM[\mKpipi][2]}. \label{eq:rmf:model:bkg}
\end{equation}

The intensity spectrum of the \WaveK4+1+\PKSt\Ppi G wave exhibits an enhanced low-mass tail below $\mKpipi\approx\SI{2}{\GeVcc}$. To take this into account we used the following modified version of \cref{eq:rmf:model:bkg} to parameterize its non-resonant component:
\begin{equation}
	\dynamicFBkg \to \frac{\dynamicFBkg}{\mathcal{K}\mKpipiTpr}. \label{eq:rmf:model:bkgNoPS}
\end{equation}

\begin{table}[!p]%
	\setTableStretchDef%
	\centering%
	\caption[Resonance components included in the \rmf.]{Resonance components included in the 14 partial waves in the RMF and listed in \cref{tab:appendix:rmf:waves}. We list the fit-parameter limits and the start-parameter ranges for the mass parameters \mO and the width parameters \GO. The resonance parameters of the \PK[1460] component were fixed in the \rmf to the values measured by LHCb~\cite{LHCb:2017swu}. The resonance parameters of the \PKOne[1400] component were fixed in the \rmf to the PDG average values~\cite{ParticleDataGroup:2024cfk}. The last column shows the decay mode that we assumed when calculating the dynamic width of the resonance in \cref{eq:pwd:model:mdepwidth}. The same decay mode is used for all resonances that belong to the same \JP sector, \ie the dominant decay channel of the dominant resonance.}%
	\label{tab:appendix:rmf:res}%
	\begin{tabular}{ll|rr|rr|r|c}
		\toprule%
		\multirow2*{Resonance} & \multirow2*{Parameter} &  \multicolumn{2}{c|}{Limits}   & \multicolumn{2}{c|}{Start Ranges} &     \multicolumn{1}{c|}{Fixed Values} & \multicolumn{1}{c}{Decay Mode} \\
		&                        & \multicolumn{2}{c|}{[\si\MeVcc]} &       \multicolumn{2}{c|}{[\si\MeVcc]}        & \multicolumn{1}{c|}{[\si\MeVcc]} & \multicolumn{1}{c}{for $\Gamma(m)$} \\ \midrule
		\multirow2*{\PK[1460]}                & \mO       &  --- &                    --- &  --- &                                 --- &      1482.4 & \multirow2*{\Decay\PKSt \Ppi P}\\
		& \GO             &  --- &                    --- &  --- &                                 --- &       335.6 &\\ \hline
		\multirow2*{\PK[1690]}                & \mO                  & 1500 &                   1700 & 1600 &                                1650 &       --- & \multirow2*{\Decay\PKSt \Ppi P}\\
		& \GO             &   10 &                    350 &  150 &                                 200 &       --- &\\ \hline
		\multirow2*{\PK[1830]}                & \mO                  & 1800 &                   1930 & 1820 &                                1860 &       --- & \multirow2*{\Decay\PKSt \Ppi P}\\
		& \GO             &   10 &                    400 &  130 &                                 200 &       --- &\\ \hline
		\multirow2*{\PKOne[1270]}             & \mO                  & 1200 &                   1500 & 1270 &                                1290 &       --- & \multirow2*{\Decay\PKSt \Ppi S}\\
		& \GO             &   50 &                    600 &   80 &                                 130 &       --- &\\ \hline
		\multirow2*{\PKOne[1400]}             & \mO                  &  --- &                    --- &  --- &                                 --- &      1403 & \multirow2*{\Decay\PKSt \Ppi S}\\
		& \GO             &  --- &                    --- &  --- &                                 --- &       174 &\\ \hline
		\multirow2*{\PKOne[1630]}             & \mO                  & 1550 &                   2300 & 1600 &                                1900 &       --- & \multirow2*{\Decay\PKSt \Ppi S}\\
		& \GO             &   50 &                    600 &  120 &                                 350 &       --- &\\ \hline
		\multirow2*{\PKTwoSt}                 & \mO                  & 1300 &                   1500 & 1425 &                                1435 &       --- & \multirow2*{\Decay K\pi D}\\
		& \GO             &   80 &                    600 &  105 &                                 115 &       --- &\\ \hline
		\multirow2*{\PKTwo[1770]}             & \mO                  & 1700 &                   1790 & 1700 &                                1790 &       --- & \multirow2*{\Decay\PKSt\Ppi P}\\
		& \GO             &  80 &                    600 &  150 &                                 200 &       --- &\\ \hline
		\multirow2*{\PKTwo[1820]}             & \mO                  & 1800 &                   2000 & 1820 &                                1850 &       --- & \multirow2*{\Decay\PKSt\Ppi P}\\
		& \GO             &  100 &                    600 &  150 &                                 250 &       --- &\\ \hline
		\multirow2*{\PKTwo[2250]}             & \mO                  & 2100 &                   2450 & 2200 &                                2280 &       --- & \multirow2*{\Decay\PKSt\Ppi P}\\
		& \GO             &   50 &                    600 &  150 &                                 250 &       --- &\\ \hline
		\multirow2*{\PKThree*}           & \mO                  & 2100 &                   2500 & 2200 &                                2450 &       --- & \multirow2*{\Decay\PKTwoSt\Ppi P}\\
		& \GO             &  100 &                    600 &  100 &                                 300 &       --- &\\ \hline
		\multirow2*{\PKFourSt}                & \mO                  & 2000 &                   2400 & 2050 &                                2080 &       --- & \multirow2*{\Decay\PKSt \Ppi G}\\
		& \GO             &  100 &                    600 &  150 &                                 250 &       --- \\  \hline
		\multirow2*{\PKFour*}            & \mO                  & 2100 &                   2650 & 2200 &                                2450 &       --- & \multirow2*{\Decay\PKSt \Ppi F}\\
		& \GO             &  100 &                    600 &  100 &                                 300 &       --- \\
		\bottomrule
		%                                     &                        &      &                        &
	\end{tabular}%
\end{table}

\clearpage
%=============================================================================
%=============================================================================
%=============================================================================
\supplementalsection{Systematic Uncertainties}

To determine the systematic uncertainties of our results for the masses and widths of the eleven strange-meson resonances we performed numerous systematic studies, which we briefly summarize in the following. Details of the performed systematic studies can be found in \refCite{Wallner:2022prx}.
These studies included the event selection, in particular the final-state particle identification. Wee varied the stringency of the final-state particle identification requirements or removed momentum regions with a potentially imperfect modeling of the final-state particle identification in the Monte Carlo detector simulation.

We performed several studies to estimate the systematic uncertainty from the partial-wave decomposition. We estimated systematic effects from the ambiguous particle identification discussed in the main text by manually removing one of the ambiguous partial waves from the set of waves included in $\sum_{a,b}^{\text{waves}}$ in equation~(1) of the main text. We also used a different approach to construct the set of waves included in $\sum_{a,b}^{\text{waves}}$ based on information field theory (see section~3 of \refCite{Kaspar:2023imx} for details).
This novel method uses a Bayesian regularization approach to narrow the large pool of allowed waves to those that significantly contribute to the data. This approach is different from the one used in the main analysis, and in addition imposes continuity of the partial waves in \mKpipi, which is not done in the main analysis.

The systematic studies also include the resonance-model fit. We studied the use of bootstrap estimates in the \chis formalism.
We also tested assumptions that enter the resonance-model fit, such as fixing the mass and width parameters of the \PKOne[1400] and \PK[1460] components, by freeing these parameters in studies. The exclusion of a second excited state \PKOnePrPr was tested by including it in a study.
We tested the choice of the 14 partial waves used in the resonance-model fit by performing resonance-model fits with various subsets of these 14 partial waves.

Finally, the systematic uncertainty for a given resonance parameter is the maximum deviation from the main analysis observed in any of the above studies.\footnote{Some of the studies performed clearly deteriorated the quality of the data or the analysis model and produced incorrect results for a limited set of resonances, particularly for small resonances. We excluded these studies when calculating the systematic uncertainties of the resonance parameters of the resonances concerned.}

\clearpage
%=============================================================================
%=============================================================================
%=============================================================================
\supplementalsection{Results}

\begin{figure*}[!t]%
	\centering%
	\subfloat[]{\includegraphics[scale=0.5]{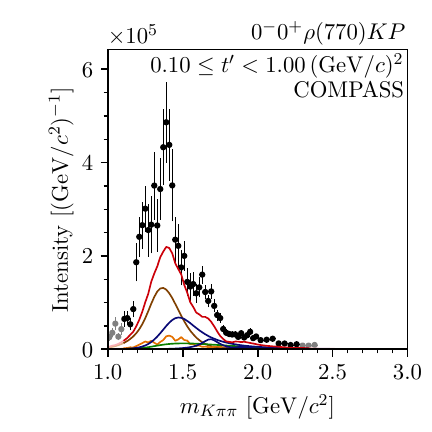}}%
	\subfloat[]{\includegraphics[scale=0.5]{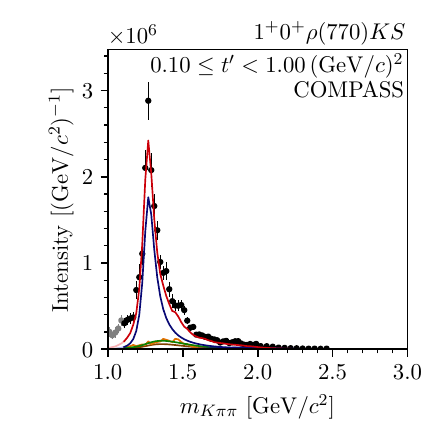}}%
	\subfloat[]{\includegraphics[scale=0.5]{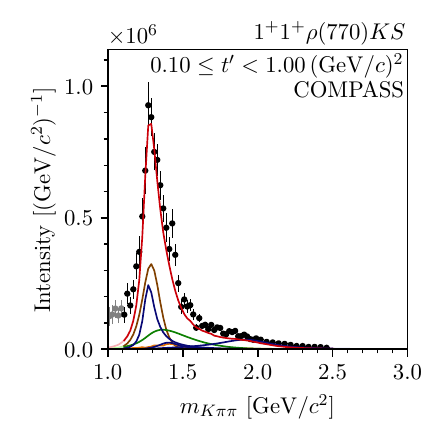}}%
	\subfloat[]{\includegraphics[scale=0.5]{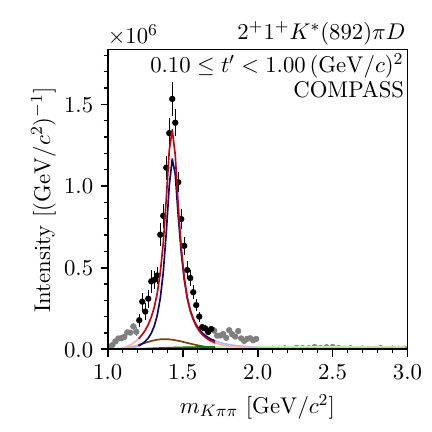}}%
	\\%
	\subfloat[]{\includegraphics[scale=0.5]{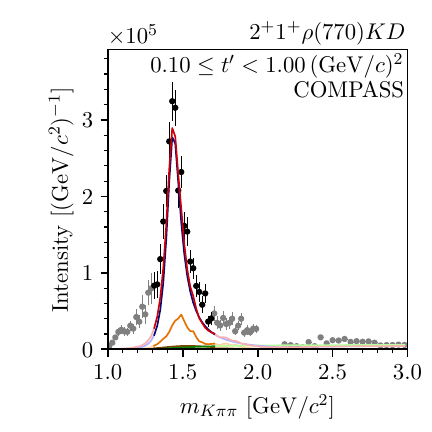}}%
	\subfloat[]{\includegraphics[scale=0.5]{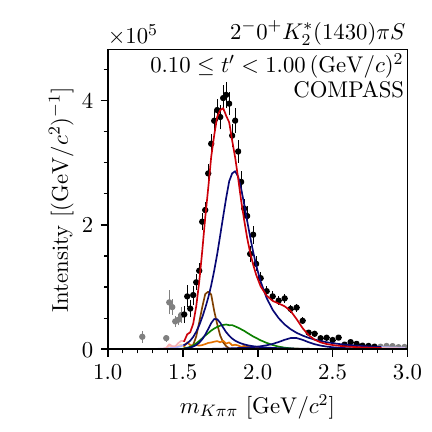}}%
	\subfloat[]{\includegraphics[scale=0.5]{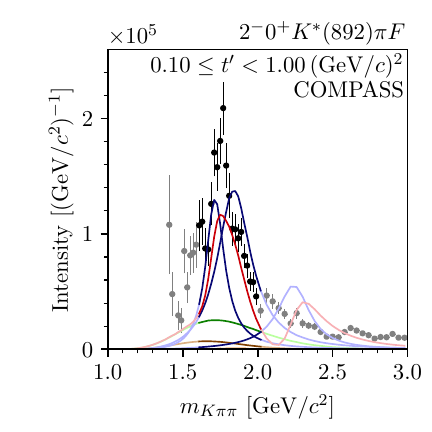}}%
	\subfloat[]{\includegraphics[scale=0.5]{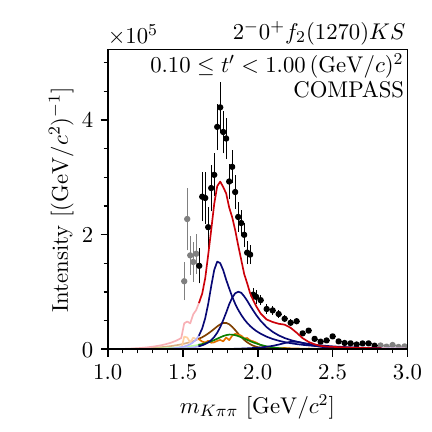}}%
	\\%
	\subfloat[]{\includegraphics[scale=0.5]{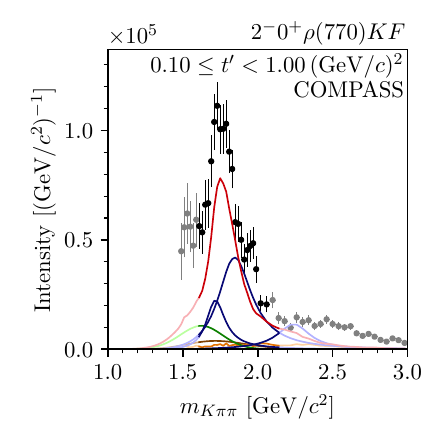}}%
	\subfloat[]{\includegraphics[scale=0.5]{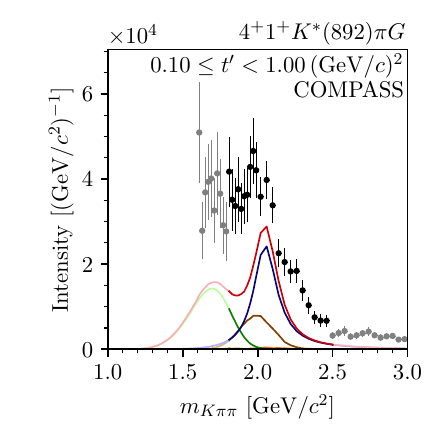}}%
	\subfloat[]{\includegraphics[scale=0.5]{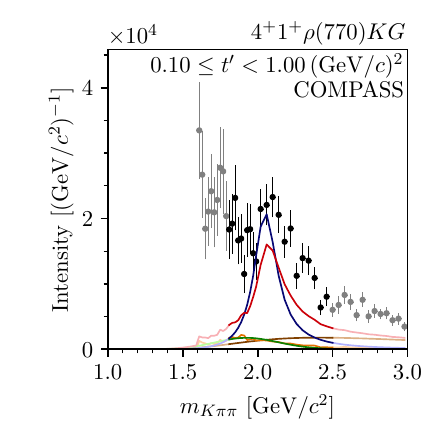}}%
	\subfloat[]{\includegraphics[scale=0.5]{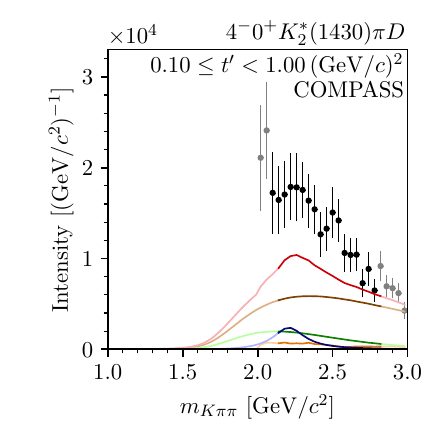}}%
	\\%
	\subfloat[]{\includegraphics[scale=0.5]{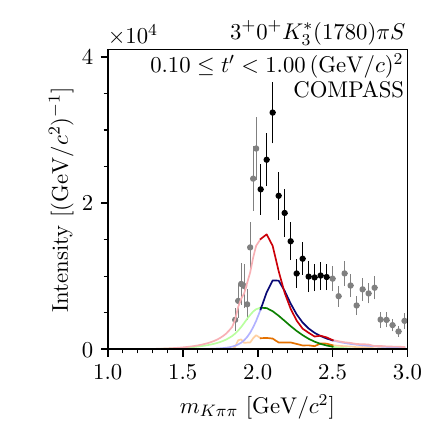}}%
	\subfloat[]{\includegraphics[scale=0.5]{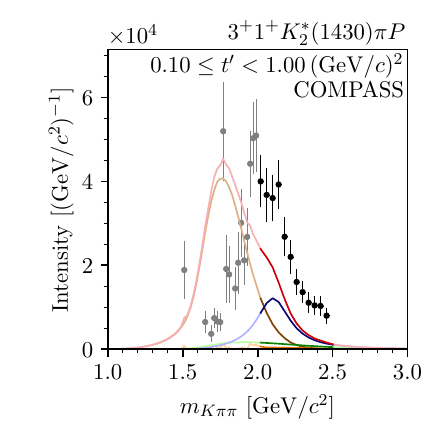}}%
	\caption{\tpr-summed intensity spectra of the 14 partial waves used in the resonance-model fit. Data points represent the measured intensities. Curves represent the \textcolor{MpRed}{total \rmfmodel}~(\textcolor{MpRed}{red}), the individual \textcolor{MpBlue}{resonance components}~(\textcolor{MpBlue}{blue}), the \textcolor{MpGreen}{non-resonant components}~(\textcolor{MpGreen}{green}), the \textcolor{MpOrange}{\threePi background components}~(\textcolor{MpOrange}{orange}), and the \textcolor{MpBrown}{\ebkg components}~(\textcolor{MpBrown}{brown}). Extrapolations of the model curves beyond fitted \mKpipi ranges are shown in lighter colors, grey for data points.}%
	\label{fig:appendix:waves}%
\end{figure*}

\begin{figure*}[!t]%
	\centering%
	\subfloat[]{\includegraphics[scale=0.5]{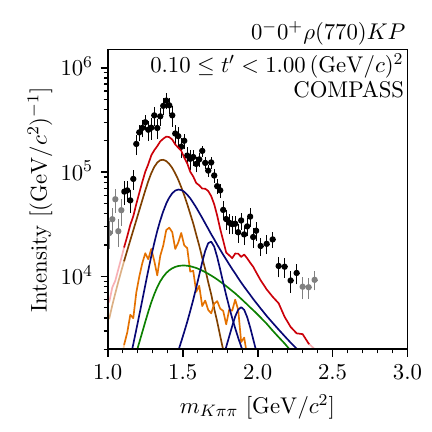}}%
	\subfloat[]{\includegraphics[scale=0.5]{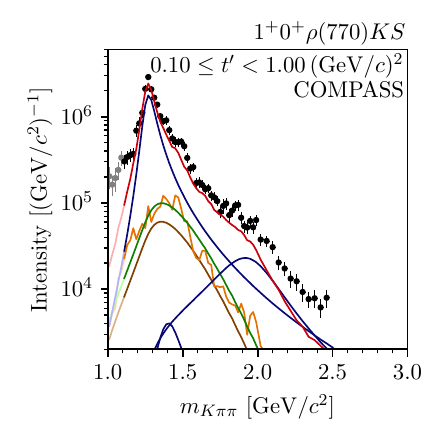}}%
	\subfloat[]{\includegraphics[scale=0.5]{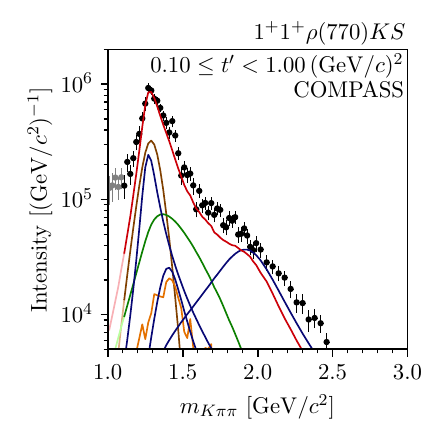}}%
	\subfloat[]{\includegraphics[scale=0.5]{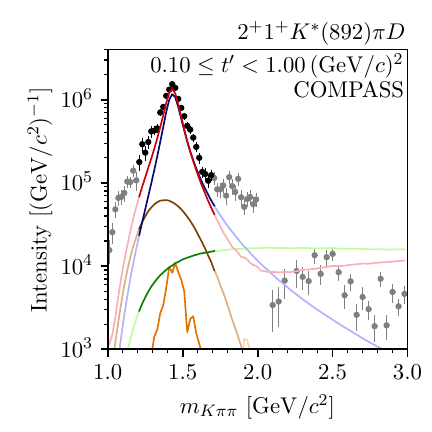}}%
	\\%
	\subfloat[]{\includegraphics[scale=0.5]{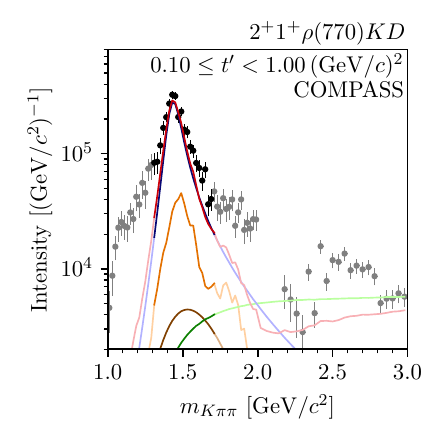}}%
	\subfloat[]{\includegraphics[scale=0.5]{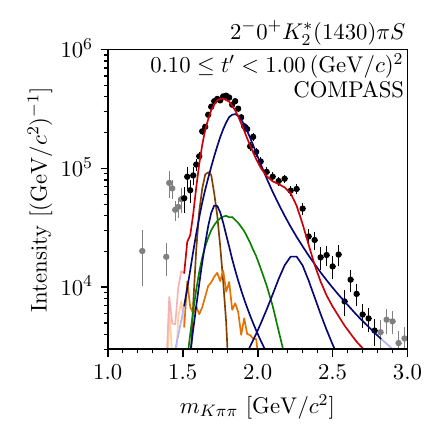}}%
	\subfloat[]{\includegraphics[scale=0.5]{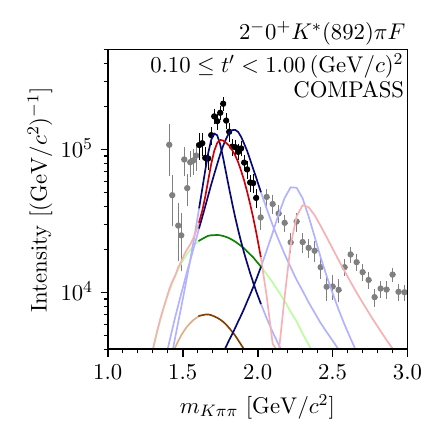}}%
	\subfloat[]{\includegraphics[scale=0.5]{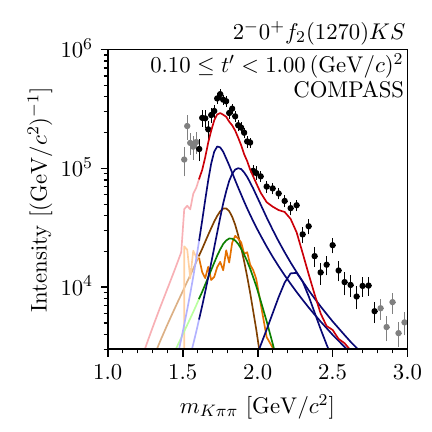}}%
	\\%
	\subfloat[]{\includegraphics[scale=0.5]{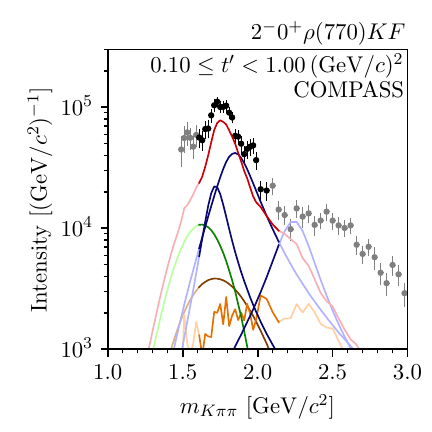}}%
	\subfloat[]{\includegraphics[scale=0.5]{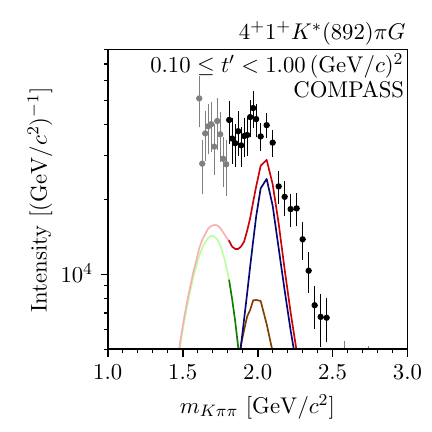}}%
	\subfloat[]{\includegraphics[scale=0.5]{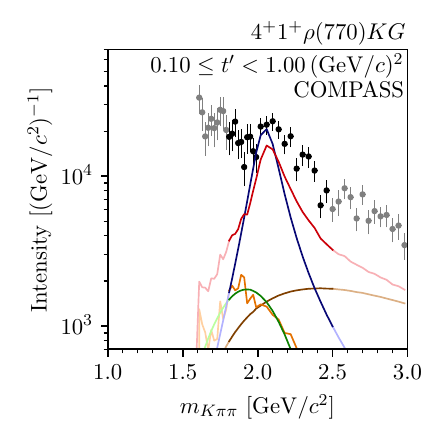}}%
	\subfloat[]{\includegraphics[scale=0.5]{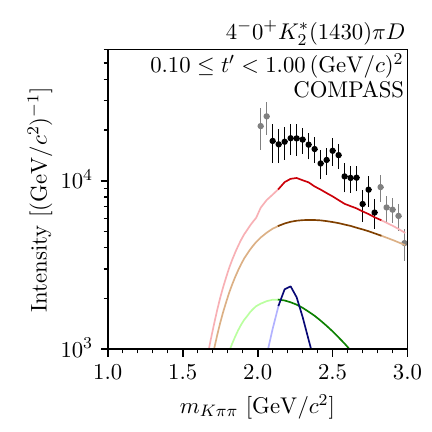}}%
	\\%
	\subfloat[]{\includegraphics[scale=0.5]{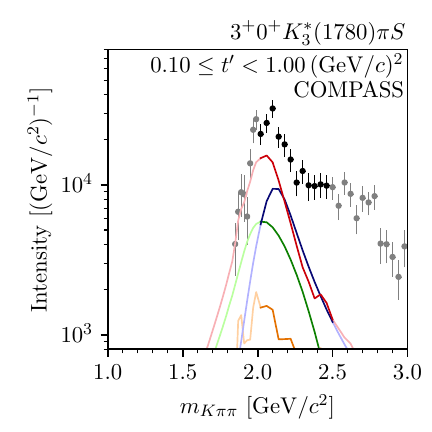}}%
	\subfloat[]{\includegraphics[scale=0.5]{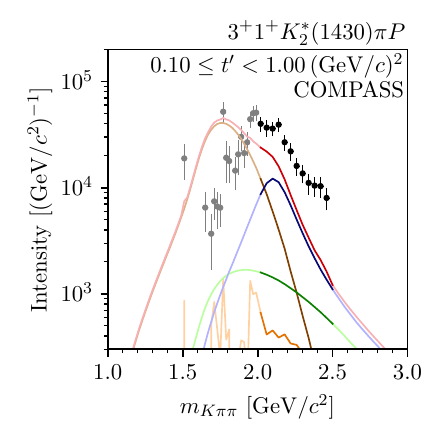}}%
	\caption{Same as \cref{fig:appendix:waves}, but showing the intensity spectra in log scale.}%
	\label{fig:appendix:waveslog}%
\end{figure*}

\begin{figure*}[!t]%
	\centering%
	\subfloat[]{\includegraphics[scale=0.5]{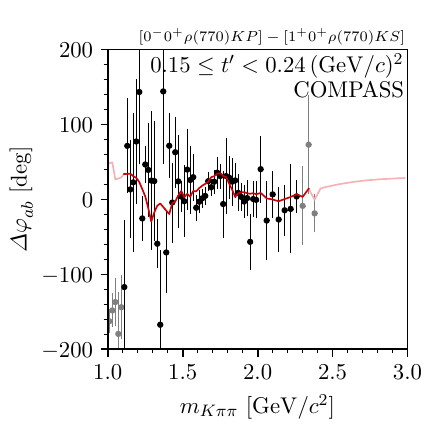}}%
	\subfloat[]{\includegraphics[scale=0.5]{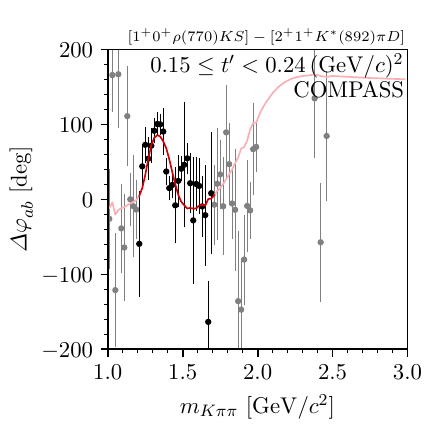}}%
	\subfloat[]{\includegraphics[scale=0.5]{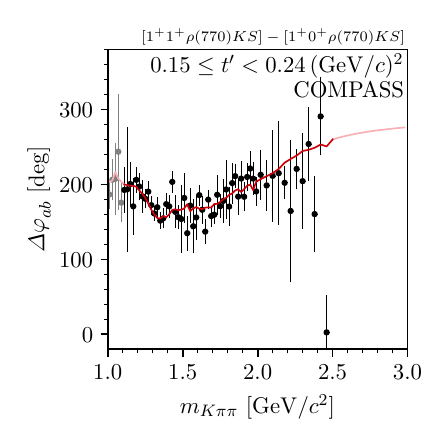}}%
	\subfloat[]{\includegraphics[scale=0.5]{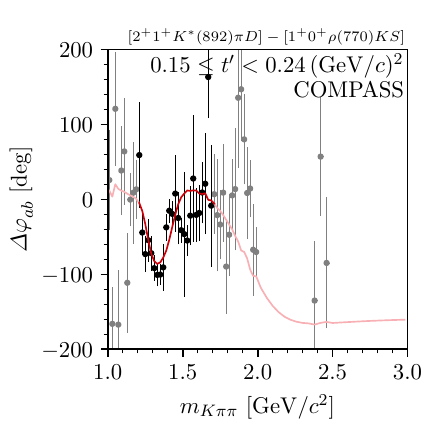}}%
	\\%
	\subfloat[]{\includegraphics[scale=0.5]{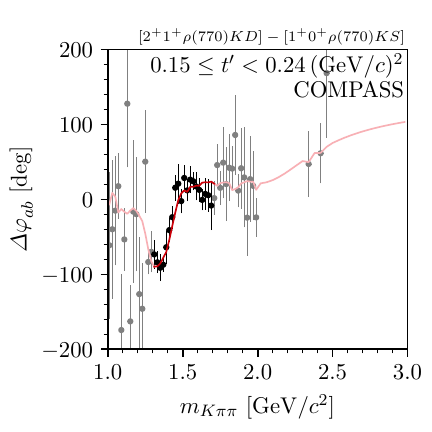}}%
	\subfloat[]{\includegraphics[scale=0.5]{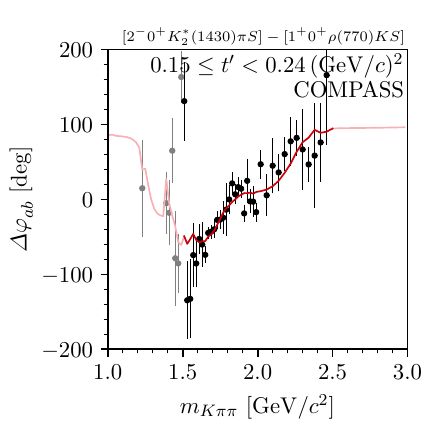}}%
	\subfloat[]{\includegraphics[scale=0.5]{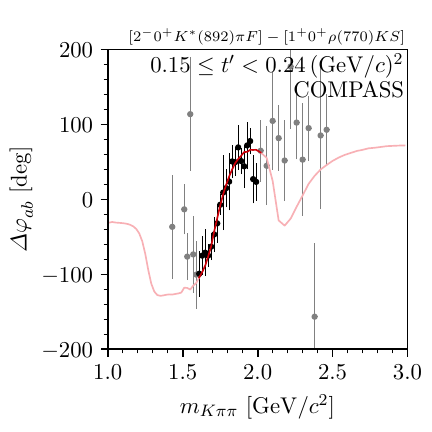}}%
	\subfloat[]{\includegraphics[scale=0.5]{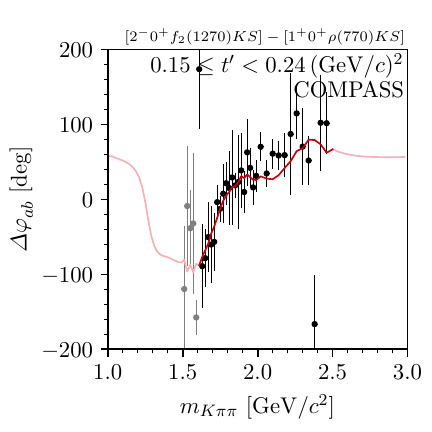}}%
	\\%
	\subfloat[]{\includegraphics[scale=0.5]{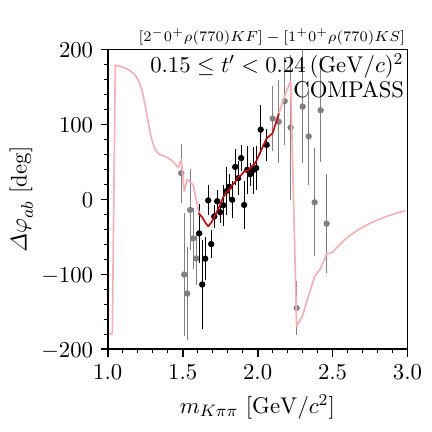}}%
	\subfloat[]{\includegraphics[scale=0.5]{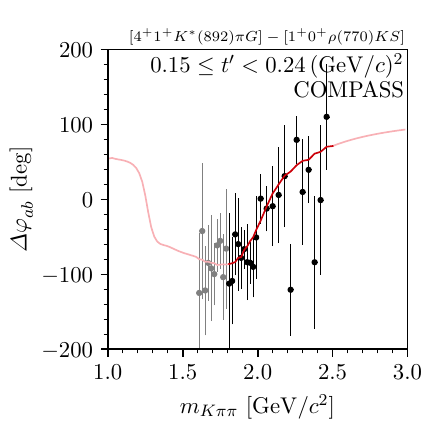}}%
	\subfloat[]{\includegraphics[scale=0.5]{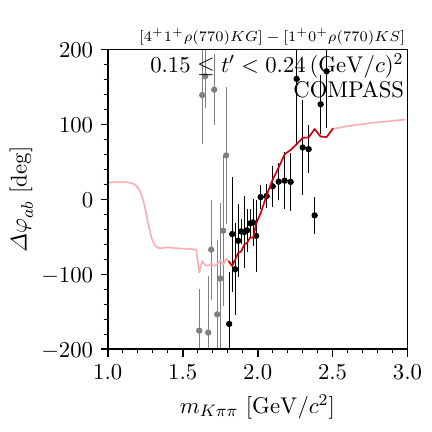}}%
	\subfloat[]{\includegraphics[scale=0.5]{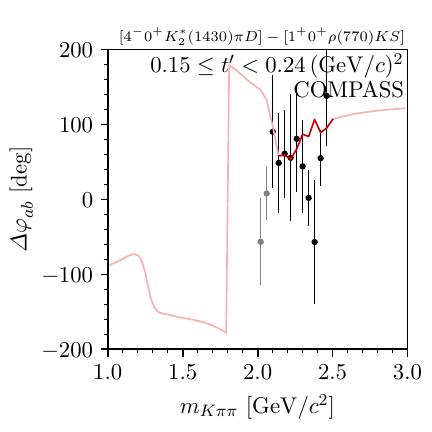}}%
	\\%
	\subfloat[]{\includegraphics[scale=0.5]{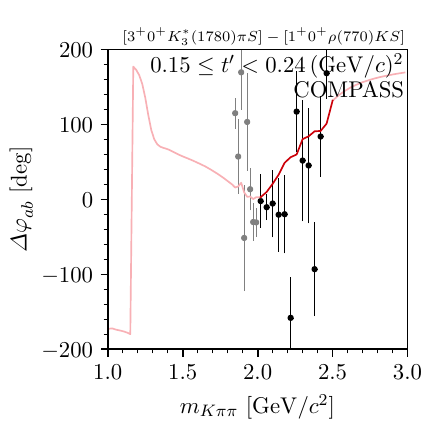}}%
	\subfloat[]{\includegraphics[scale=0.5]{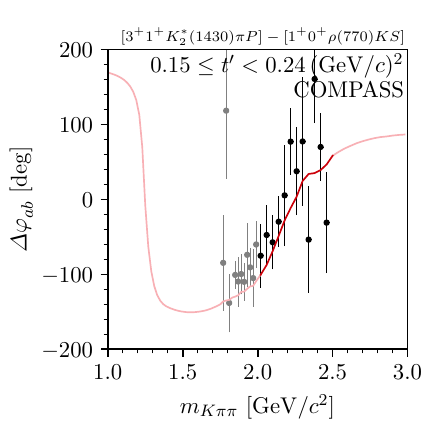}}%
	\caption{Phases of the 14 partial waves used in the resonance-model fit in the second-lowest \tpr bin relative to the \WaveK1+0+\Prho\PK S wave, except for the phase of the \WaveK1+0+\Prho\PK S wave, which is shown relative to the \WaveK2+1+ \PKSt\Ppi D wave. Same color code as in \cref{fig:appendix:waves}.}%
	\label{fig:appendix:phases}%
\end{figure*}

\begin{figure*}[!t]
	\centering%
	\includegraphics[width=\linewidth]{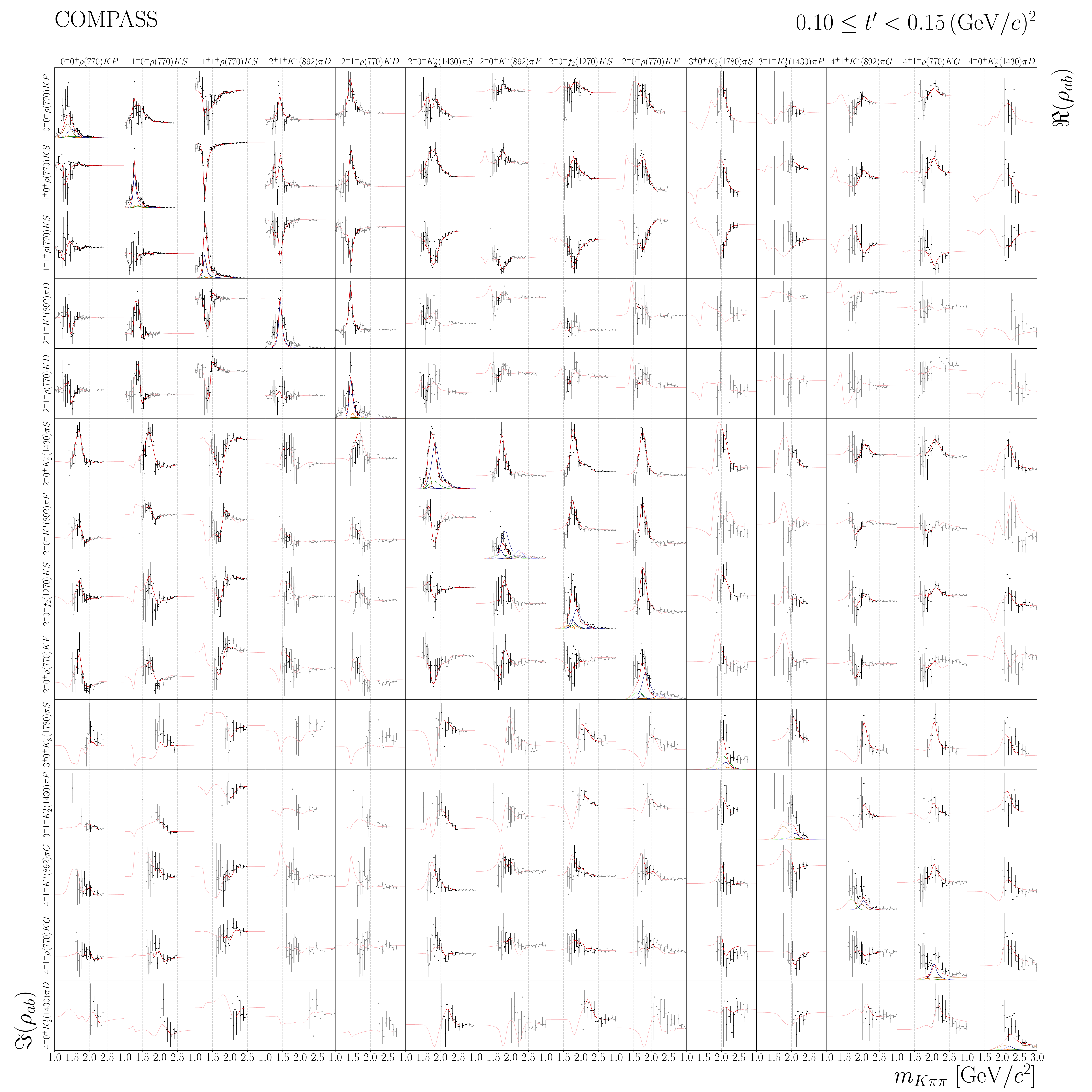}%
	\caption{(Zoom-in version for online view) Real and imaginary parts of the spin-density matrix elements as a function of \mKpipi in the lowest of the four \tpr bins for the 14 partial waves that were included in the resonance-model fit. The graphs on the diagonal show the intensity spectra. The upper-right and lower-left off-diagonal graphs show the real and imaginary parts, respectively, of the off-diagonal elements of the spin-density matrix. Same color code as in \cref{fig:appendix:waves}.}
	\label{fig:appendix:sdm:t1}
\end{figure*}

\begin{figure*}[!t]
	\centering%
	\includegraphics[width=\linewidth]{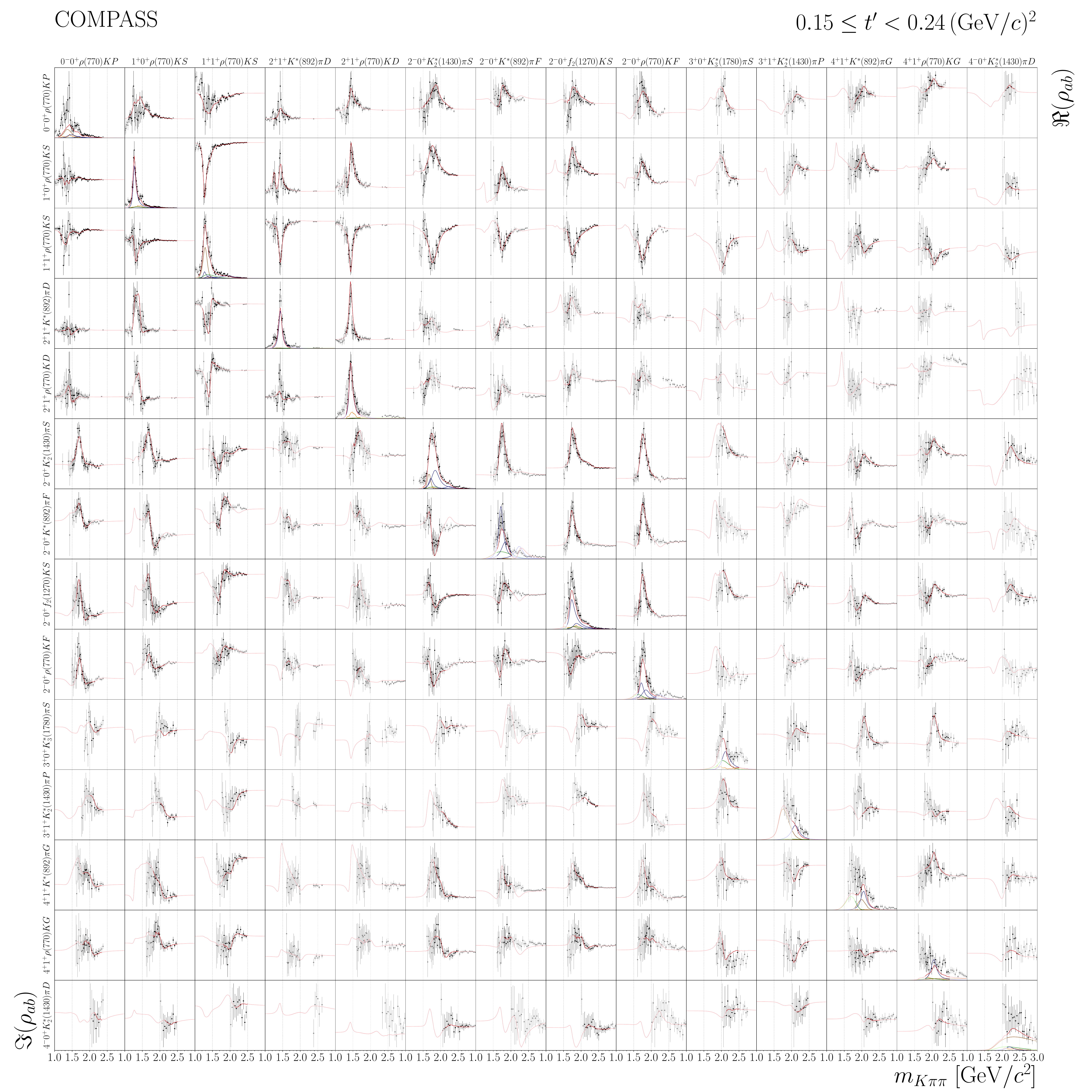}%
	\caption{Same as \cref{fig:appendix:sdm:t1}, but showing the spin-density matrix elements in the second-lowest \tpr bin.}
	\label{fig:appendix:sdm:t2}
\end{figure*}

\begin{figure*}[!t]
	\centering%
	\includegraphics[width=\linewidth]{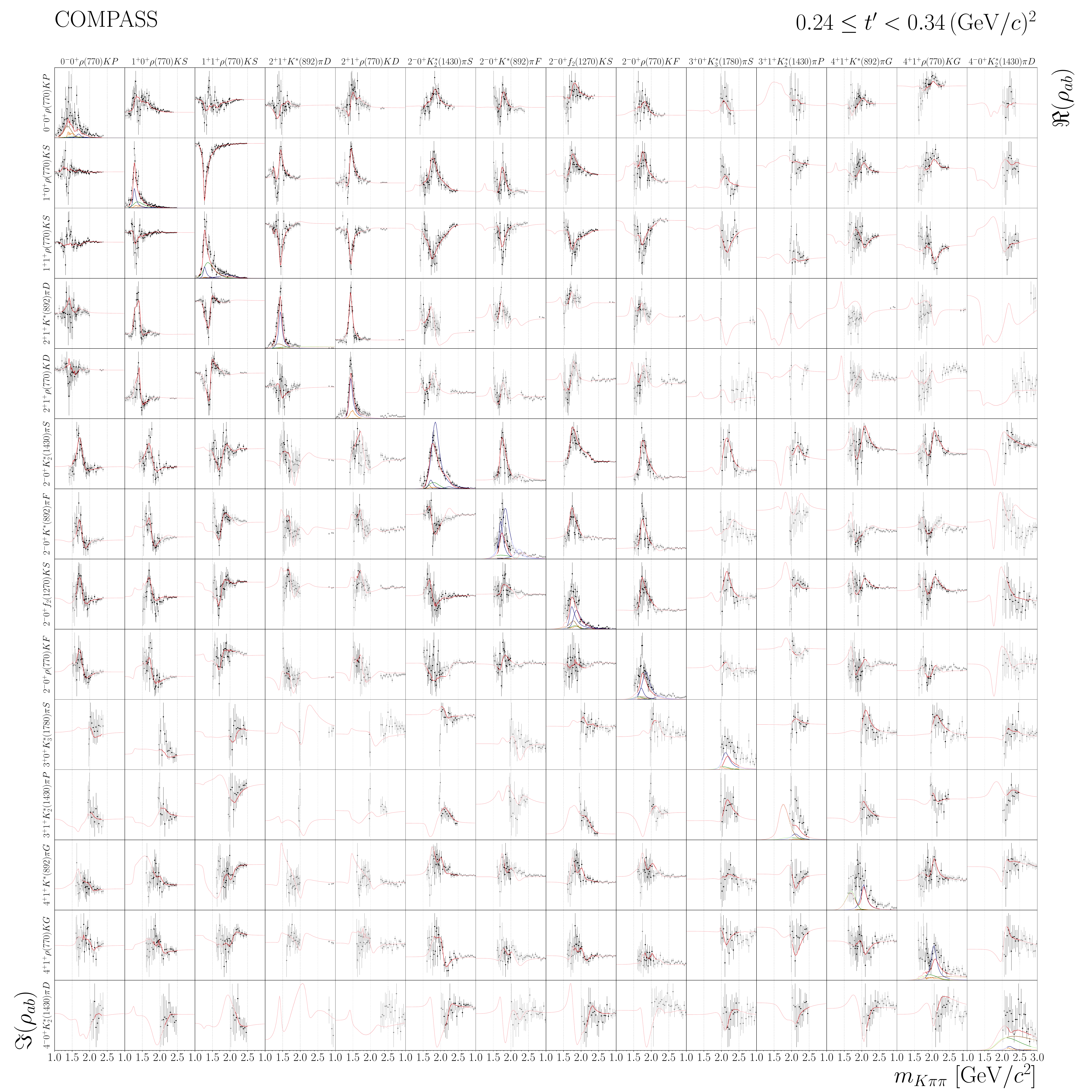}%
	\caption{Same as \cref{fig:appendix:sdm:t1}, but showing the spin-density matrix elements in the third-lowest \tpr bin.}
	\label{fig:appendix:sdm:t3}
\end{figure*}

\begin{figure*}[!t]
	\centering%
	\includegraphics[width=\linewidth]{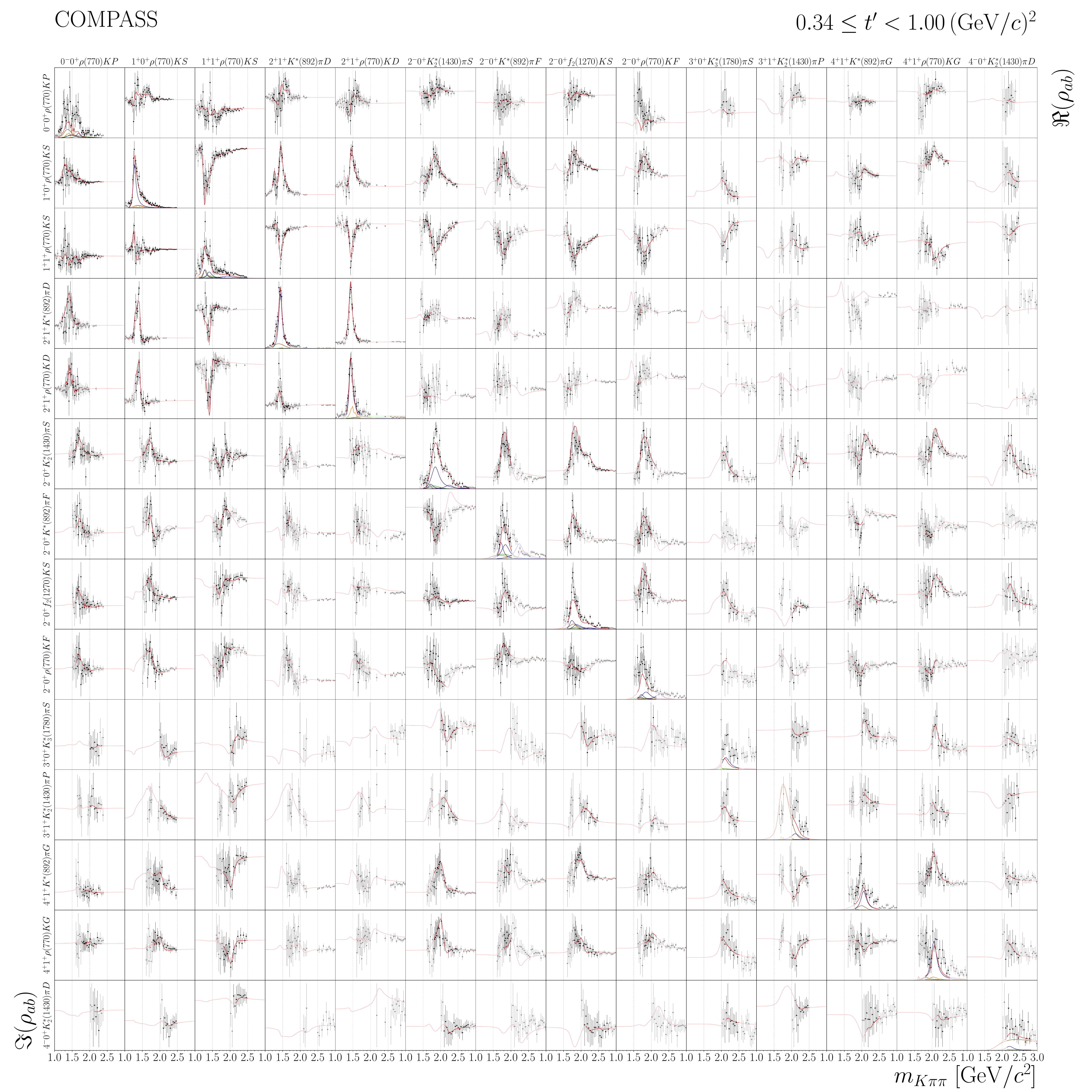}%
	\caption{Same as \cref{fig:appendix:sdm:t1}, but showing the spin-density matrix elements in the highest \tpr bin.}
	\label{fig:appendix:sdm:t4}
\end{figure*}

This section summarizes the results of the partial-wave analysis of the reaction \reactionK. \Cref{tab:appendix:results:resparametersI} lists our results for the resonance parameters of the eleven strange-meson resonances together with the corresponding values from PDG or from individual previous measurements (see table caption for details).
\Cref{fig:appendix:waves,fig:appendix:waveslog} show the intensity spectra of the 14 partial waves used in the resonance-model fit in linear and logarithmic scale, respectively. \Cref{fig:appendix:phases} shows the phases of the 14 partial waves used in the resonance-model fit in the second \tpr bin relative to the \WaveK1+0+\Prho\PK S wave.
\Cref{fig:appendix:sdm:t1,fig:appendix:sdm:t2,fig:appendix:sdm:t3,fig:appendix:sdm:t4} show the real and imaginary parts of the spin-density matrix elements of the 14 partial waves used in the resonance-model fit, \ie the data that enters the resonance-model fit, and the corresponding model curves obtained from the resonance-model fit in the four \tpr bins.

\begin{table}[!p]%
	\setTableStretchDef%
	\centering%
	\captionabove{Resonance parameters as obtained from the 14-wave \rmf. The first quoted uncertainties are statistical, the second systematic uncertainties. The values and uncertainties are rounded to the same precision according to the PDG rounding rules~\cite{ParticleDataGroup:2024cfk}. The number of significant digits is given by the total uncertainty. For the total uncertainty, we quadratically add the statistical uncertainty to the upper and lower systematic uncertainties. For comparison, the PDG averages from \refCite{ParticleDataGroup:2024cfk} are listed. The PDG lists more than one average value for the \PKTwoSt resonance. We list here the PDG average values for the charged \PKTwoSt decaying into the $\PK\Ppi$ final state. The PDG does not quote averages for the parameters of the \PK[1630], \PK[1830], and \PKFour[2500] as there is only a single measurement for each state considered by the PDG. We quote the masses and widths from these measurement~\cite{Karnaukhov:1998qq,LHCb:2016axx,Cleland:1980ya}.}%
	\label{tab:appendix:results:resparametersI}%
	\subfloat[$K$-like resonances]{
		\centering
		\begin{tabular}{cl|ll}
			%\toprule 
			&                 & \multicolumn{1}{c}{\PK[1690]/\PK[1630]}                           & \multicolumn{1}{c} {\PK[1830]}     \\
			\midrule
			\multirow{2}*{\includegraphics[width=0.83cm]{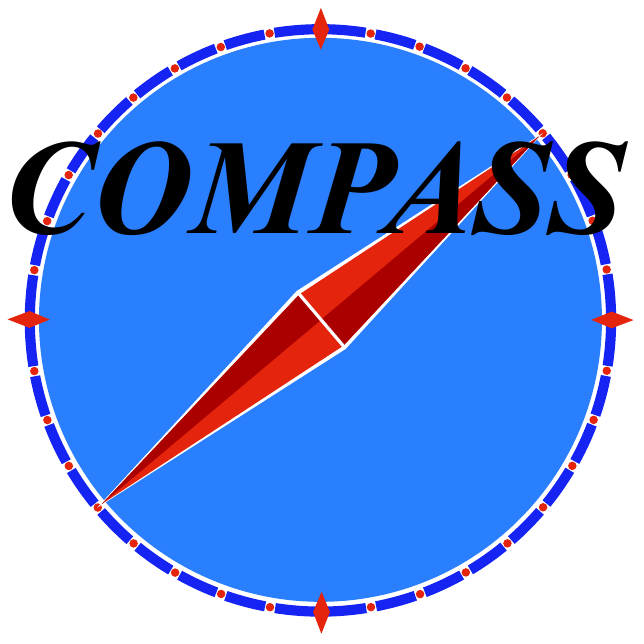}} & \mO [\si\MeVcc] & \MK*{1630}                                               & \MK*{1830}                         \\
			& \GO [\si\MeVcc] & \GK*{1630}                                               & \GK*{1830}              \\
			\hline
			\multirow{2}*{\rotatebox[origin=c]{90}{PDG}}               & \mO [\si\MeVcc] & ${1629}\valueSep\pm\phantom{1} 7$                                   & $\numerrsa[parse-numbers=false]{1874}{43}{+\phantom{1}59}{-115}$ \\
			& \GO [\si\MeVcc] & $\phantom{12}\numaerr[parse-numbers=false]{16}{+19}{-16}$& $\phantom{1}\numerrsa[parse-numbers=false]{168}{90}{+208}{-104}$ \\
			\bottomrule
		\end{tabular}%
	}
	
	\vspace*{6ex}%
	\subfloat[$K_1$-like resonances]{
		\centering
		\begin{tabular}{cl|ll}
			%\toprule
			&                 & \multicolumn{1}{c}{\PKOne[1270]}                     & \multicolumn{1}{c} {\PKOnePr}     \\
			\midrule
			\multirow{2}*{\includegraphics[width=0.83cm]{logos/compass.pdf}} & \mO [\si\MeVcc] & \MKOne*{1270}                                        & \MKOne*{1650}                         \\
			& \GO [\si\MeVcc] & \GKOne*{1270}                                        & \GKOne*{1650}              \\
			\hline
			\multirow{2}*{\rotatebox[origin=c]{90}{PDG}}               & \mO [\si\MeVcc] & ${1253}\valueSep\pm 7$        & $1650 \valueSep\pm 50$                \\
			& \GO [\si\MeVcc] & $\phantom{12}90\valueSep\pm20$ & $\phantom{1}150\valueSep\pm 50$       \\
			\bottomrule
		\end{tabular}%
	}
	
	\vspace*{6ex}%
	\subfloat[$K_2$-like resonances]{
		\centering
		\begin{tabular}{cl|lll}
			%\toprule
			&                 & \multicolumn{1}{c}{\PKTwo[1770]}             & \multicolumn{1}{c}{\PKTwo[1820]}      & \multicolumn{1}{c} {\PKTwo[2250]}     \\
			\midrule
			\multirow{2}*{\includegraphics[width=0.83cm]{logos/compass.pdf}} & \mO [\si\MeVcc] & \MKTwo*{1770}                                & \MKTwo*{1820}                         & \MKTwo*{2250}                         \\
			& \GO [\si\MeVcc] & \GKTwo*{1770}                                & \GKTwo*{1820}              & \GKTwo*{2250}              \\
			\hline
			\multirow{2}*{\rotatebox[origin=c]{90}{PDG}}               & \mO [\si\MeVcc] & $1773\valueSep\pm \phantom{} 8$            & $1819\valueSep\pm 12$                 & $2247\valueSep\pm  17$                \\
			& \GO [\si\MeVcc] & $\phantom{1}186\valueSep\pm  14$ & $\phantom{1}264\valueSep\pm 34$      & $\phantom{1}180 \valueSep\pm  30$     \\
			\bottomrule
		\end{tabular}%
	}
\end{table}

\begin{table}[!p]%
	\ContinuedFloat%
	\setTableStretchDef%
	\centering%
	\captionabove{Continued.}%
	\setcounter{subtable}{3}%
	\label{tab:appendix:results:resparametersII}
	\subfloat[$K_3$- and $K_4$-like resonances]{
		\centering
		\begin{tabular}{cl|ll}
			%\toprule
			&                 & \multicolumn{1}{c}{\PKThree*}                   & \multicolumn{1}{c} {\PKFour*}     \\
			\midrule
			\multirow{2}*{\includegraphics[width=0.83cm]{logos/compass.pdf}} & \mO [\si\MeVcc] & \MKThree*{2320}                                      & \MKFour*{2500}                         \\
			& \GO [\si\MeVcc] & \GKThree*{2320}                                      & \GKFour*{2500}              \\
			\hline
			\multirow{2}*{\rotatebox[origin=c]{90}{PDG}}               & \mO [\si\MeVcc] & ${2324}\valueSep\pm 24$                              & $2490 \valueSep\pm 20$                \\
			& \GO [\si\MeVcc] & $\phantom{1}150\valueSep\pm30$                       & $\approx 250$                         \\
			\bottomrule
		\end{tabular}%
	}
	
	\vspace*{6ex}%
	\subfloat[$K^{*}_J$-like resonances]{
		\centering
		\begin{tabular}{cl|ll}
			%\toprule
			&                 & \multicolumn{1}{c}{\PKTwoSt}          & \multicolumn{1}{c}{\PKFourSt}                                     \\
			\midrule
			\multirow{2}*{\includegraphics[width=0.83cm]{logos/compass.pdf}} & \mO [\si\MeVcc] & \MKTwoSt*{1430}                       & \MKFourSt*{2045}                                                  \\
			& \GO [\si\MeVcc] & \GKTwoSt*{1430}                       & \GKFourSt*{2045}                                       \\
			\hline
			\multirow{2}*{\rotatebox[origin=c]{90}{PDG}}               & \mO [\si\MeVcc] & $1427.3 \valueSep\pm  1.5$            & \numaerr[parse-numbers=false]{2048}{+\phantom{1}8}{-\phantom{1}9} \\
			& \GO [\si\MeVcc] & \phantom{1}$100.0 \valueSep\pm  2.1$  & $\phantom{2}\numaerr{199}{27}{-19}$                               \\
			\bottomrule
		\end{tabular}%
	}	
	
\end{table}
\FloatBarrier

\clearpage

%\bibliography{Refs.bib}
%apsrev4-2.bst 2019-01-14 (MD) hand-edited version of apsrev4-1.bst
%Control: key (0)
%Control: author (72) initials jnrlst
%Control: editor formatted (1) identically to author
%Control: production of article title (-1) disabled
%Control: page (0) single
%Control: year (1) truncated
%Control: production of eprint (0) enabled
%

%\printbibliography[heading=bibintoc]%
\clearpage
\center{\textbf{The COMPASS Collaboration}}

\vspace{10pt}
\begin{flushleft}
G.~D.~Alexeev$^\textrm{{\footnotesize\hyperlink{hl:dubna}{28}}}$\orcidlink{0009-0007-0196-8178},
M.~G.~Alexeev$^\textrm{{\footnotesize\hyperlink{hl:turin_u}{20},\hyperlink{hl:turin_i}{19}}}$\orcidlink{0000-0002-7306-8255},
C.~Alice$^\textrm{{\footnotesize\hyperlink{hl:turin_u}{20},\hyperlink{hl:turin_i}{19}}}$\orcidlink{0000-0001-6297-9857},
A.~Amoroso$^\textrm{{\footnotesize\hyperlink{hl:turin_u}{20},\hyperlink{hl:turin_i}{19}}}$\orcidlink{0000-0002-3095-8610},
V.~Andrieux$^\textrm{{\footnotesize\hyperlink{hl:illinois}{33}}}$\orcidlink{0000-0001-9957-9910},
V.~Anosov$^\textrm{{\footnotesize\hyperlink{hl:dubna}{28}}}$\orcidlink{0009-0003-3595-9561},
K.~Augsten$^\textrm{{\footnotesize\hyperlink{hl:praguectu}{4}}}$\orcidlink{0000-0001-8324-0576},
W.~Augustyniak$^\textrm{{\footnotesize\hyperlink{hl:warsaw}{23}}}$,
C.~D.~R.~Azevedo$^\textrm{{\footnotesize\hyperlink{hl:aveiro}{26}}}$\orcidlink{0000-0002-0012-9918},
B.~Badelek$^\textrm{{\footnotesize\hyperlink{hl:warsawu}{25}}}$\orcidlink{0000-0002-4082-1466},
R.~Beck$^\textrm{{\footnotesize\hyperlink{hl:bonniskp}{8}}}$,
J.~Beckers$^\textrm{{\footnotesize\hyperlink{hl:munichtu}{12}}}$\orcidlink{0009-0009-7186-255X},
Y.~Bedfer$^\textrm{{\footnotesize\hyperlink{hl:saclay}{6}}}$\orcidlink{0000-0002-5198-1852},
J.~Bernhard$^\textrm{{\footnotesize\hyperlink{hl:cern}{30}}}$\orcidlink{0000-0001-9256-971X},
F.~Bradamante$^\textrm{{\footnotesize\hyperlink{hl:triest_i}{17}}}$\orcidlink{0000-0001-6136-376X},
A.~Bressan$^\textrm{{\footnotesize\hyperlink{hl:triest_u}{18},\hyperlink{hl:triest_i}{17}}}$\orcidlink{0000-0002-3718-6377},
W.-C.~Chang$^\textrm{{\footnotesize\hyperlink{hl:taipei}{31}}}$\orcidlink{0000-0002-1695-7830},
C.~Chatterjee$^\textrm{{\footnotesize\hyperlink{hl:triest_i}{17},\hyperlink{hl:a}{a}}}$\orcidlink{0000-0001-7784-3792},
M.~Chiosso$^\textrm{{\footnotesize\hyperlink{hl:turin_u}{20},\hyperlink{hl:turin_i}{19}}}$\orcidlink{0000-0001-6994-8551},
S.-U.~Chung$^\textrm{{\footnotesize\hyperlink{hl:munichtu}{12},\hyperlink{hl:j}{j},\hyperlink{hl:j1}{j1}}}$,
A.~Cicuttin$^\textrm{{\footnotesize\hyperlink{hl:triest_i}{17},\hyperlink{hl:triest_a}{16}}}$\orcidlink{0000-0002-3645-9791},
M.~L.~Crespo$^\textrm{{\footnotesize\hyperlink{hl:triest_i}{17},\hyperlink{hl:triest_a}{16}}}$\orcidlink{0000-0002-5483-3388}
% ...

\vspace{10pt}
\hypertarget{hl:aanl}{$^\textrm{{\footnotesize 1}}$\footnotesize~A.I. Alikhanyan National Science Laboratory, 2 Alikhanyan Br. Street, 0036, Yerevan, Armenia$^\textrm{{\tiny\hyperlink{hl:A}{A}}}$\\}
\hypertarget{hl:brno}{$^\textrm{{\footnotesize 2}}$\footnotesize~Institute of Scientific Instruments of the CAS, 61264 Brno, Czech Republic$^\textrm{{\tiny\hyperlink{hl:B}{B}}}$\\}
\hypertarget{hl:liberec}{$^\textrm{{\footnotesize 3}}$\footnotesize~Technical University in Liberec, 46117 Liberec, Czech Republic$^\textrm{{\tiny\hyperlink{hl:B}{B}}}$\\}
\hypertarget{hl:praguectu}{$^\textrm{{\footnotesize 4}}$\footnotesize~Czech Technical University in Prague, 16636 Prague, Czech Republic$^\textrm{{\tiny\hyperlink{hl:B}{B}}}$\\}
\hypertarget{hl:praguecu}{$^\textrm{{\footnotesize 5}}$\footnotesize~Charles University, Faculty of Mathematics and Physics, 12116 Prague, Czech Republic$^\textrm{{\tiny\hyperlink{hl:B}{B}}}$\\}
\hypertarget{hl:saclay}{$^\textrm{{\footnotesize 6}}$\footnotesize~IRFU, CEA, Universit\'e Paris-Saclay, 91191 Gif-sur-Yvette, France\\}
\hypertarget{hl:bochum}{$^\textrm{{\footnotesize 7}}$\footnotesize~Universit\"at Bochum, Institut f\"ur Experimentalphysik, 44780 Bochum, Germany$^\textrm{{\tiny\hyperlink{hl:C}{C}}}$\\}
\hypertarget{hl:bonniskp}{$^\textrm{{\footnotesize 8}}$\footnotesize~Universit\"at Bonn, Helmholtz-Institut f\"ur  Strahlen- und Kernphysik, 53115 Bonn, Germany$^\textrm{{\tiny\hyperlink{hl:C}{C}}}$\\}
\hypertarget{hl:bonnpi}{$^\textrm{{\footnotesize 9}}$\footnotesize~Universit\"at Bonn, Physikalisches Institut, 53115 Bonn, Germany$^\textrm{{\tiny\hyperlink{hl:C}{C}}}$\\}
\hypertarget{hl:freiburg}{$^\textrm{{\footnotesize 10}}$\footnotesize~Universit\"at Freiburg, Physikalisches Institut, 79104 Freiburg, Germany$^\textrm{{\tiny\hyperlink{hl:C}{C}}}$\\}
\hypertarget{hl:mainz}{$^\textrm{{\footnotesize 11}}$\footnotesize~Universit\"at Mainz, Institut f\"ur Kernphysik, 55099 Mainz, Germany$^\textrm{{\tiny\hyperlink{hl:C}{C}}}$\\}
\hypertarget{hl:munichtu}{$^\textrm{{\footnotesize 12}}$\footnotesize~Technische Universit\"at M\"unchen, Physik Dept., 85748 Garching, Germany$^\textrm{{\tiny\hyperlink{hl:C}{C}}}$\\}
\hypertarget{hl:munichuni}{$^\textrm{{\footnotesize 13}}$\footnotesize~Ludwig-Maximilians-Universit\"at, 80539 M\"unchen, Germany\\}
\hypertarget{hl:calcutta}{$^\textrm{{\footnotesize 14}}$\footnotesize~Matrivani Institute of Experimental Research \& Education, Calcutta-700 030, India$^\textrm{{\tiny\hyperlink{hl:D}{D}}}$\\}
\hypertarget{hl:telaviv}{$^\textrm{{\footnotesize 15}}$\footnotesize~Tel Aviv University, School of Physics and Astronomy, 69978 Tel Aviv, Israel$^\textrm{{\tiny\hyperlink{hl:E}{E}}}$\\}
\hypertarget{hl:triest_a}{$^\textrm{{\footnotesize 16}}$\footnotesize~Abdus Salam ICTP, 34151 Trieste, Italy\\}
\hypertarget{hl:triest_i}{$^\textrm{{\footnotesize 17}}$\footnotesize~Trieste Section of INFN, 34127 Trieste, Italy\\}
\hypertarget{hl:triest_u}{$^\textrm{{\footnotesize 18}}$\footnotesize~University of Trieste, Dept.\ of Physics, 34127 Trieste, Italy\\}
\hypertarget{hl:turin_i}{$^\textrm{{\footnotesize 19}}$\footnotesize~Torino Section of INFN, 10125 Torino, Italy\\}
\hypertarget{hl:turin_u}{$^\textrm{{\footnotesize 20}}$\footnotesize~University of Torino, Dept.\ of Physics, 10125 Torino, Italy\\}
\hypertarget{hl:miyazaki}{$^\textrm{{\footnotesize 21}}$\footnotesize~University of Miyazaki, Miyazaki 889-2192, Japan$^\textrm{{\tiny\hyperlink{hl:F}{F}}}$\\}
\hypertarget{hl:yamagata}{$^\textrm{{\footnotesize 22}}$\footnotesize~Yamagata University, Yamagata 992-8510, Japan$^\textrm{{\tiny\hyperlink{hl:F}{F}}}$\\}
\hypertarget{hl:warsaw}{$^\textrm{{\footnotesize 23}}$\footnotesize~National Centre for Nuclear Research, 02-093 Warsaw, Poland$^\textrm{{\tiny\hyperlink{hl:G}{G}}}$\\}
\hypertarget{hl:warsawtu}{$^\textrm{{\footnotesize 24}}$\footnotesize~Warsaw University of Technology, Institute of Radioelectronics, 00-665 Warsaw, Poland$^\textrm{{\tiny\hyperlink{hl:G}{G}}}$\\}
\hypertarget{hl:warsawu}{$^\textrm{{\footnotesize 25}}$\footnotesize~University of Warsaw, Faculty of Physics, 02-093 Warsaw, Poland$^\textrm{{\tiny\hyperlink{hl:G}{G}}}$\\}
\hypertarget{hl:aveiro}{$^\textrm{{\footnotesize 26}}$\footnotesize~University of Aveiro, I3N, Dept. of Physics, 3810-193 Aveiro, Portugal$^\textrm{{\tiny\hyperlink{hl:H}{H}}}$\\}
\hypertarget{hl:lisbon}{$^\textrm{{\footnotesize 27}}$\footnotesize~LIP, 1649-003 Lisbon, Portugal$^\textrm{{\tiny\hyperlink{hl:H}{H}}}$\\}
\hypertarget{hl:dubna}{$^\textrm{{\footnotesize 28}}$\footnotesize~Affiliated with an international laboratory covered by a cooperation agreement with CERN\\}
\hypertarget{hl:russia}{$^\textrm{{\footnotesize 29}}$\footnotesize~Affiliated with an institute formerly covered by a cooperation agreement with CERN\\}
\hypertarget{hl:cern}{$^\textrm{{\footnotesize 30}}$\footnotesize~CERN, 1211 Geneva 23, Switzerland\\}
\hypertarget{hl:taipei}{$^\textrm{{\footnotesize 31}}$\footnotesize~Academia Sinica, Institute of Physics, Taipei 11529, Taiwan$^\textrm{{\tiny\hyperlink{hl:I}{I}}}$\\}
\hypertarget{hl:taipeincu}{$^\textrm{{\footnotesize 32}}$\footnotesize~Center for High Energy and High Field Physics and Dept.\ of Physics, National Central University, 300 Zhongda Rd., Zhongli 320317, Taiwan$^\textrm{{\tiny\hyperlink{hl:I}{I}}}$\\}
\hypertarget{hl:illinois}{$^\textrm{{\footnotesize 33}}$\footnotesize~University of Illinois at Urbana-Champaign, Dept.\ of Physics, Urbana, IL 61801-3080, USA$^\textrm{{\tiny\hyperlink{hl:J}{J}}}$\\}

\vspace{10pt}
\hypertarget{hl:*}{$^\textrm{{\footnotesize *}}$\footnotesize~Corresponding author\\}
\hypertarget{hl:a}{$^\textrm{{\footnotesize a}}$\footnotesize~Supported by the European Union’s Horizon 2020 research and innovation programme under grant agreement STRONG–2020 - No 824093\\}
\hypertarget{hl:b}{$^\textrm{{\footnotesize b}}$\footnotesize~Retired from Ludwig-Maximilians-Universit\"at, 80539 M\"unchen, Germany\\}
\hypertarget{hl:b1}{$^\textrm{{\footnotesize b1}}$\footnotesize~Supported by the DFG cluster of excellence `Origin and Structure of the Universe' (www.universe-cluster.de) (Germany)\\}
\hypertarget{hl:c}{$^\textrm{{\footnotesize c}}$\footnotesize~Also at ORIGINS Excellence Cluster, 85748 Garching, Germany\\}
\hypertarget{hl:d}{$^\textrm{{\footnotesize d}}$\footnotesize~Also at Institut f\"ur Theoretische Physik, Universit\"at T\"ubingen, 72076 T\"ubingen, Germany\\}
\hypertarget{hl:e}{$^\textrm{{\footnotesize e}}$\footnotesize~Supported by the Max Planck Institute for Physics, 85748 Garching, Germany\\}
\hypertarget{hl:f}{$^\textrm{{\footnotesize f}}$\footnotesize~Present address: NISER, Centre for Medical and Radiation Physics, Bubaneswar, India\\}
\hypertarget{hl:g}{$^\textrm{{\footnotesize g}}$\footnotesize~Also at University of Eastern Piedmont, 15100 Alessandria, Italy\\}
\hypertarget{hl:g1}{$^\textrm{{\footnotesize g1}}$\footnotesize~Supported by the Funds for Research 2019-22 of the University of Eastern Piedmont\\}
\hypertarget{hl:h}{$^\textrm{{\footnotesize h}}$\footnotesize~Also at Chubu University, Kasugai, Aichi 487-8501, Japan\\}
\hypertarget{hl:i}{$^\textrm{{\footnotesize i}}$\footnotesize~Also at KEK, 1-1 Oho, Tsukuba, Ibaraki 305-0801, Japan\\}
\hypertarget{hl:j}{$^\textrm{{\footnotesize j}}$\footnotesize~Also at Dept.\ of Physics, Pusan National University, Busan 609-735, Republic of Korea\\}
\hypertarget{hl:j1}{$^\textrm{{\footnotesize j1}}$\footnotesize~Also at Physics Dept., Brookhaven National Laboratory, Upton, NY 11973, USA\\}
\hypertarget{hl:k}{$^\textrm{{\footnotesize k}}$\footnotesize~Also at Fairmont State University, Department of Natural Sciences, 1201 Locust Ave, Fairmont, West Virginia 26554, USA\\}
\hypertarget{hl:l}{$^\textrm{{\footnotesize l}}$\footnotesize~Also at Dept.\ of Physics, National Kaohsiung Normal University, Kaohsiung County 824, Taiwan\\}
\hypertarget{hl:m}{$^\textrm{{\footnotesize m}}$\footnotesize~Also at RIKEN Nishina Center for Accelerator-Based Science, Wako, Saitama 351-0198, Japan\\}
\hypertarget{hl:$\dagger$}{$^\textrm{{\footnotesize $\dagger$}}$\footnotesize~Deceased\\}

\vspace{10pt}
\hypertarget{hl:A}{$^\textrm{{\tiny A}}$\footnotesize~Supported by the Higher Education and Science Committee of the Republic of Armenia (Armenia)\\}
\hypertarget{hl:B}{$^\textrm{{\tiny B}}$\footnotesize~Supported by MEYS, Grants LM2023040, LM2018104, LTT17018 and GAUK60121, CZ.02.01.01/00/22\_008/0004632 "FORTE", co-funded by the EU and Charles University Grant PRIMUS/22/SCI/017 (Czech Republic)\\}
\hypertarget{hl:C}{$^\textrm{{\tiny C}}$\footnotesize~Supported by BMBF - Bundesministerium f\"ur Bildung und Forschung (Germany)\\}
\hypertarget{hl:D}{$^\textrm{{\tiny D}}$\footnotesize~Supported by B. Sen fund (India)\\}
\hypertarget{hl:E}{$^\textrm{{\tiny E}}$\footnotesize~Supported by the Israel Academy of Sciences and Humanities (Israel)\\}
\hypertarget{hl:F}{$^\textrm{{\tiny F}}$\footnotesize~Supported by MEXT and JSPS, Grants 18002006, 20540299, 18540281 and 26247032, the Daiko and Yamada Foundations (Japan)\\}
\hypertarget{hl:G}{$^\textrm{{\tiny G}}$\footnotesize~Supported by NCN, Grant 2020/37/B/ST2/01547 (Poland)\\}
\hypertarget{hl:H}{$^\textrm{{\tiny H}}$\footnotesize~Supported by FCT, Grants DOI 10.54499/CERN/FIS-PAR/0022/2019 and DOI 10.54499/CERN/FIS-PAR/0016/2021 (Portugal)\\}
\hypertarget{hl:I}{$^\textrm{{\tiny I}}$\footnotesize~Supported by the Ministry of Science and Technology (Taiwan)\\}
\hypertarget{hl:J}{$^\textrm{{\tiny J}}$\footnotesize~Supported by the National Science Foundation, Grant no. PHY-1506416 (USA)\\}

\end{flushleft}

\end{document}